\documentclass[12pt,authoryear]{article}
\usepackage[paperheight=11.7in,paperwidth=8.3in,margin=1in]{geometry}

\usepackage[utf8]{inputenc}
\usepackage{amsmath}
\usepackage{amssymb}
\usepackage{amsfonts}
\usepackage{bm}

\usepackage{appendix}
\usepackage{enumerate}
\usepackage{enumitem}
\usepackage{nicefrac}
\usepackage{array}
\usepackage{calc}
\usepackage{flafter}
\usepackage{esdiff}
\usepackage{mathabx}
\usepackage{textcomp}
\usepackage{color}		
\usepackage{graphicx}
\usepackage{amscd}
\usepackage{mathrsfs}
\usepackage{enumerate}
\usepackage{mdwlist}
\usepackage[round]{natbib}
\usepackage{authblk}
\usepackage[colorlinks=true,linkcolor=blue,citecolor=blue,urlcolor=blue]{hyperref}
\usepackage[export]{adjustbox}

\usepackage{algorithm}
\usepackage[noend]{algpseudocode}

\usepackage{caption}
\usepackage{subcaption}

\hypersetup{
    colorlinks,
    citecolor=black,
    filecolor=black,
    linkcolor=black,
    urlcolor=black
}

\makeatletter
\def\BState{\State\hskip-\ALG@thistlm}
\makeatother

\makeatletter
\def\step{%
   \@ifnextchar[ \@myitem{\@noitemargtrue\@myitem[\@itemlabel]}}
\def\@myitem[#1]{\item[#1]\mbox{}}
\makeatother

\newcommand\R{\mathbb{R}}

\newcommand\E{\mathbb{E}}

\newcommand\grad{\nabla}

\newcommand{\calF}{\mathcal{F}}

\newcommand{\calM}{\mathcal{M}}

\newcommand{\calT}{\mathcal{T}}

\newcommand{\calV}{\mathcal{V}}
\newcommand{\calW}{\mathcal{W}}

\newcommand{\calZ}{\mathcal{Z}}
\newcommand{\ra}{\rightarrow}
\newcommand{\la}{\leftarrow}

\renewcommand\epsilon{\varepsilon}

\newcommand{\vect}[1]{\mathbf{#1}}
\newcommand{\Id}{\text{Id}}

\DeclareMathOperator*{\argmin}{arg\,min}

\newcommand{\Mtemp}{{\calM_{0}}}
\newcommand{\MtempT}{{\calM_{0}^{\calT}}}

\newcommand{\M}{\calM}
\newcommand{\Diff}{\text{Diff}}

\newcommand{\FoS}{FoS}
\newcommand{\FoSs}{FoSs}

\graphicspath{{../}}

\title{Statistical Analysis of Functions on Surfaces, with an application to Medical Imaging}
\author[1]{Eardi Lila\thanks{e.lila@maths.cam.ac.uk}}
\author[2]{John A. D. Aston\thanks{j.aston@statslab.cam.ac.uk}}
\affil[1]{Cambridge Centre for Analysis, University of Cambridge}
\affil[2]{Statistical Laboratory, DPMMS, University of Cambridge}
\date{\today}

\begin{document}

\maketitle

\begin{abstract}
In Functional Data Analysis, data are commonly assumed to be smooth functions on a fixed interval of the real line. In this work, we introduce a comprehensive framework for the analysis of functional data, whose domain is a two-dimensional manifold and the domain itself is subject to variability from sample to sample. We formulate a statistical model for such data, here called Functions on Surfaces, which enables a joint representation of the geometric and functional aspects, and propose an associated estimation framework. We assess the validity of the framework by performing a simulation study and we finally apply it to the analysis of neuroimaging data of cortical thickness, acquired from the brains of different subjects, and thus lying on domains with different geometries.
\end{abstract}

\section{Introduction}\label{sec:intro}
\begin{figure}[!htb]
\centering
\includegraphics[width=1\textwidth]{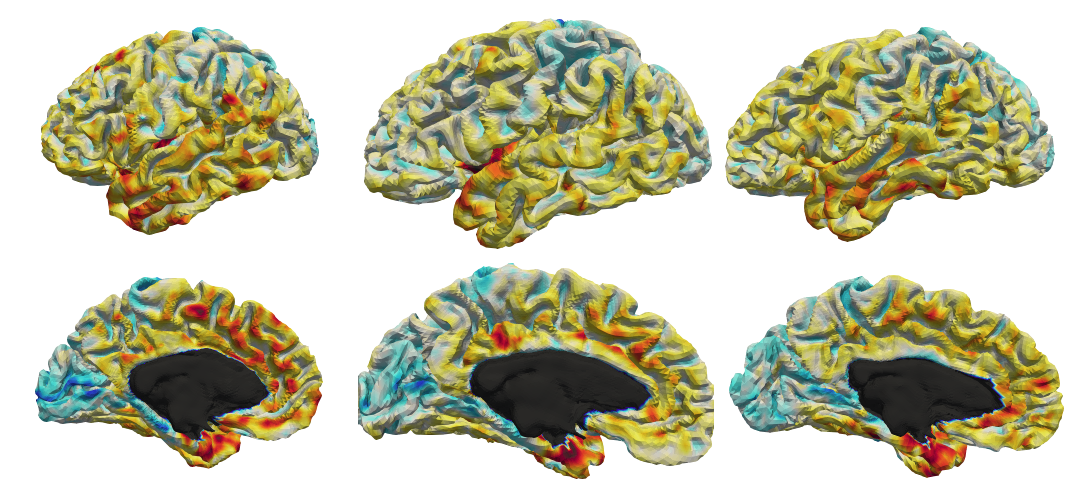}
\caption[]{Surface reconstructions of the brain's left hemispheres of three different subjects, with an associated scalar signal representing the cerebral cortex thickness of the subjects. These have been reconstructed from 3D MRI scans of the subjects. The black area is a region which is not part of the brain surface.}
\label{fig:cortical_surfaces}
\end{figure}

Advances in medical imaging acquisition are constantly increasing the complexity of data representing anatomical objects. In particular, some of these imaging modalities offer a richer representation of anatomical manifolds, as a geometric object coupled with a function defined on the geometric object itself, i.e. a Function on a Surface (\FoS). In this work we focus on Functions on Surfaces (\FoSs) that are real functions located on domains that are two-dimensional manifolds, where the domains themselves are subject to variability from sample to sample, as shown in Figure~\ref{fig:cortical_surfaces}. In the applied mathematics literature, these are also known with the name of Functional Shapes \citep{Charon2014}. However, as it will be clear from the methodological section of this paper, the proposed framework can be extended to deal with more complex situations, such as vector-valued functions describing features arising from multi-modal imaging techniques or the RGB representation of colors, as done in \cite{Yao2017}, with the purpose of inferring the underlying geometry. Further extensions could also include situations where the functions have an inherent time component. For simplicity of exposition, we will concentrate on univariate FoS data in this paper.

The aim of the present paper is the introduction of a comprehensive statistical framework for the analysis of \FoSs. To this end, a statistical model is introduced, with the main aim of jointly representing the \textit{geometric variability} and \textit{functional variability} of the data. Suppose there is an underlying true one-to-one correspondence between the points on the geometries of the observed \FoSs. By geometric variability we mean variations on the shape of the domains, i.e. variations of the point positions from one \FoS\ to another.  By functional variability we mean variations on the amplitude of the functions of the observed \FoSs, at the points in correspondence. For instance, it is evident that the three \FoSs\ in Figure~\ref{fig:cortical_surfaces}, show both geometric variability and functional variability. In order to quantify these two types of variability, we introduce estimators of the underlying unknown quantities within the proposed statistical model.

In this context, estimating functional variability is challenging because, in such high-dimensional settings, there is need to incorporate prior information, like smoothness, on complex domains. Estimating geometric variability is challenging because the space where the geometric objects live is non-Euclidean, and this invalidates classical linear models, which could lead to predictions that do not belong to the original space, for instance, self-intersecting objects. The formulation of estimators constrained to lie in the deformation space is therefore required. Moreover, as it is clear from their definitions, the study of geometric variability cannot be performed independently from the study of functional variability, as the results of the latter generally depend on the former. This motivates the introduction of a novel diffeomorphic registration algorithm for functional data whose domain is a two-dimensional manifold, which enables the exploitation of the functional information to achieve a better registration.

Often times practitioners have approached the analysis of \FoSs\ in two completely separate steps. In the first step, the surfaces are registered to a template surface, and the functions are transported on the template through such estimated registration maps. In the second step, the analysis of the functions is performed on the template surface, independently of the previous step. This approach has two main drawbacks. Firstly, the complete separation of the two steps precludes any study aimed at understanding how the geometric variability relates to the functional variability. Secondly, for each subject, there is an infinite number of registrations that bring the template to match the target surface. However, for different registration maps, the registered functions exhibit different functional variability. In other words, the registration step is responsible for separating the variability due to geometric differences from the variability due to differences in the functions, and this strongly influences the subsequent analysis on the functions. Thus, the two steps should not be performed independently. Indeed, in the one-dimensional analogue situation, it has been seen that considerably more information can be gleaned from a joint approach than a step-wise approach \cite[see][and references therein]{marron2015}.

Many ideas from the literature on image registration \citep[see e.g.][]{Thirion1995, Thirion1998, Dupuis1998} and the literature on landmarked shapes \citep[see e.g.][]{Bookstein1997,Bookstein1997a} have been recently extended to the more general setting of surfaces, without a functional component, both from the applied mathematics prospective \citep{Younes2010} and from the statistical prospective \citep{Patrangenaru2015}. It is also natural to contextualize \FoSs\ in the Functional Data Analysis (FDA) framework. However, FDA is generally performed in controlled environments, where data are assumed to be smooth functions on a fixed interval of the real line \citep{Ramsay2005}, or more generally, smooth functions on a fixed domain. The setting considered here represents a new challenge for this branch of statistics.

More recently, a joint mathematical model for geometric and functional variability has been proposed in \cite{Charlier2017}. The approach consists of generalizing the notion of deformation to a notion of metamorphosis, introduced for 2D images in \cite{Trouve2005}. A metamorphosis includes both a geometric deformation term and an additive functional term. This enables the representation of any \FoS\ as a metamorphosis of a template \FoS. The geometric deformation and the functional additive term, to explain a given \FoS, can be weighted by two different parameters in the model. In contrast, our approach takes a statistical perspective on the problem of analyzing a set of FoSs, and aims to offer a methodological toolset that can be feasibly applied to the analysis of the brain surfaces shown in Figure~\ref{fig:cortical_surfaces}.

\subsection{Motivating application}
\begin{figure}[!htb]
\centering
\includegraphics[width=\textwidth]{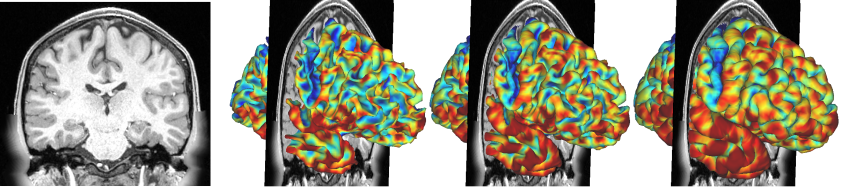}
\caption[]{From left to right, in the first panel we have a section of a structural MRI of a single subject. In the next three panels we have, respectively, the estimated inner, mid-thickness and outer surfaces of the cerebral cortex. The inner and outer surfaces enclose the cerebral cortex, the mid-thickness surface interpolates its middle points and it is used to represent its geometry. These images have been produced using Connectome Workbench \citep{glasser2013}.}
\label{fig:HCP_extraction}
\end{figure}

The motivating application of the proposed model is the study of a collection of \FoSs\ derived from Magnetic Resonance Imaging (MRI). A 2D surface representing the geometry of the cerebral cortex, the outermost layer of the brain, can be extracted from 3D MRI data thanks to fully automated surface-extraction algorithms \citep{glasser2013}. The cerebral cortex is a highly convoluted thin sheet of 2 to 4 millimeters of thickness which consists of neuronal cell bodies and it is the source of large parts of our neuronal activity. An illustration of the surface-extraction step is shown in Figure~\ref{fig:HCP_extraction}.

Thanks to complementary imaging techniques, like functional MRI, a function can be associated to the estimated cerebral cortex \citep[see, e.g.,][]{Hagler2006}, resulting in a \FoS. Such functions can be vector-valued functions, where each component represents a feature of the cerebral cortex, extracted from a different imaging technique. However, in this work, the function we consider is the map of thickness measurements of the cerebral cortex. In fact, thanks to the recent improvements of the resolution of MRI scans it is now possible to have an accurate estimation of this thickness map \citep{Lerch2005}. Details on the cerebral cortex surface reconstruction and the cortical thickness estimation are covered in Section~\ref{sec:application}. In Figure~\ref{fig:cortical_surfaces} we show the \FoSs\ representing cerebral cortex geometry and thickness of three different individuals.

Almost all studies of these kind presume a preprocessing registration step and so do not consider the inherent variability effects that might be induced by the registration step on the functional measurements. Indeed, this issue goes beyond neuroimaging, as the same techniques are often used in a wide variety of medical imaging settings \citep{Audette2000}, as well as computer vision applications \citep{Zaetz2015}.

The rest of the article is organized as follows. In Section~\ref{sec:methodology} we introduce a generative statistical model which allows for both geometric and functional variability. In Section~\ref{sec:estimation_framework}, we propose the statistical estimators of the underlying unknown quantities of the generative model. We perform a simulation study on synthetic data in Section~\ref{sec:simulations}, to investigate our estimation procedure. We then apply the framework introduced to study the relation between geometry and thickness of the human cerebral cortex in Section~\ref{sec:application} and draw some concluding remarks in Section~\ref{sec:conclusions}. Moreover, in the appendices, we present further details of the proposed methodology and an additional simulation study.


\section{Model for Functions on Surfaces}\label{sec:methodology}
\subsection{Definitions}\label{sec:methodology_def}
A set of \FoSs, such as the ones in Figure~\ref{fig:cortical_surfaces}, can be mathematically formulated as a collection of pairs $\{(\M_i, Y_i): i=1,\ldots,n\}$. The collection $\{\M_i: i=1,\ldots,n\}$ is a set of topologically equivalent smooth two-dimensional manifolds, embedded in $\R^3$, representing the geometry of the data. The functional aspect of the data is represented by the collection $\{Y_i: i=1,\ldots,n\}$, where $Y_i$ is an element of the function space $L^2(\M_i)$, i.e. the Hilbert space of square integrable functions on $\M_i$ with respect to the area measure.

Here, we propose a statistical generative model for \FoSs, modelled in terms of mathematically more tractable objects. To this end, we define a \textit{deformation operator} $\varphi$, such that $\varphi_v: \R^3 \ra \R^3$ is parametrized by the elements of a Hilbert space $\{v : v \in\calV\}$. Moreover, we assume $\varphi_v$ is a homeomorphism of $\R^3$ for all $v \in \calV$ and that $\varphi_0(x) = x$ for all $ x \in \R^3$. For each $v \in \calV$, $\varphi_v: \R^3 \ra \R^3$ represents a deformation of the space $\R^3$, which means that when $\varphi_v$ is applied to a point $x \in \R^3$ this is relocated to the location $\varphi_v(x) \in \R^3$. In addition, $\varphi_v$ being a homeomorphism of $\R^3$ implies that, for a fixed $v \in \calV$, there is a one-to-one correspondence between each element $x \in \R^3$ and the relocated element $\varphi_v(x) \in \R^3$.

Moreover, we introduce $\Mtemp$, a smooth two-dimensional manifold topologically equivalent to $\{\M_i\}$, which represents a fixed template geometric object. Given a \FoS, the geometric template together with the deformation operator offers an alternative representation of the geometry of the \FoS\ in hand as: $\varphi_v \circ \Mtemp$, for a particular choice of $v \in \calV$. Here, $\varphi_v \circ \Mtemp$ is the geometric object obtained by deforming $\Mtemp$ through the map $\varphi_v$, and specifically, by relocating each point $ x \in \Mtemp$ to the new location $\varphi_v(x)$, to resemble the target manifold. For this reason, we will informally say that the element $v \in \calV$ encodes the geometry, or the shape, of a \FoS, as in fact $v$ defines the deformation $\varphi_v$, which defines the geometry $\varphi_v \circ \Mtemp$. The choice of the deformation operator is driven by the particular problem in hand. We first introduce the generative model and subsequently discuss different choices of this operator.

\subsection{The model}
Let now $\{v_i: i=1,\ldots,n\}$ be a set of random samples of a zero-mean and finite second moment $\calV$-valued random function $V$ and $\{Z_i: i=1,\ldots,n\}$ be a set of random smooth samples of a zero-mean and finite second moment random real function $Z$ with values in $L^2(\Mtemp)$. We assume the following generative model for the $i$th observation $(\M_i, Y_i)$:
\begin{align}\label{eq:generation}
\begin{cases}
\M_i &= \varphi_{v_i} \circ \Mtemp,\\
X_i &= \mu + \delta Z_i,\\
Y_i &= X_i \circ \varphi^{-1}_{v_i},
\end{cases}
\end{align}
where $\mu \in L^2(\Mtemp)$ is a fixed function, modelling the common function behavior between the different samples, and $\delta$ is a coefficient representing the magnitude of the function variations around the mean $\mu$. In addition, we assume the objects in Model~\ref{eq:generation} are subject to a discretization error, which is considered in the estimation process. This formulation generalizes an often used model for the one-dimensional functional registration problem (see, e.g. \citet{Tang2008}).

Model~\ref{eq:generation} achieves the goal of representing \FoSs\ as a collection of more tractable objects, decomposing the generation of the $i$th \FoS\ into three main steps. In the first step, the geometry $\M_i$ of the $i$th object is generated by the deformation $\varphi_{v_i}$ applied to the template $\Mtemp$, where $v_i$ is a random sample of $V$. In the second step, a random function $X_i$, on the template, is generated as the sum of the fixed function $\mu$ and a stochastic term $\delta Z_i$. In the third step the generated function $X_i$ is transported on the manifold $\M_i$, defining $Y_i$. This is done through the equation $Y_i = X_i \circ \varphi^{-1}_{v_i}$, which means that for all $x \in \Mtemp$, $Y_i(\varphi_{v_i}(x)) = X_i(x)$, or informally that the functional value $X_i(x)$ is `transported' with the deformation to the location $\varphi_{v_i}(x) \in \M_i$.

\begin{figure}[!htb]
\centering
\includegraphics[width=1\textwidth]{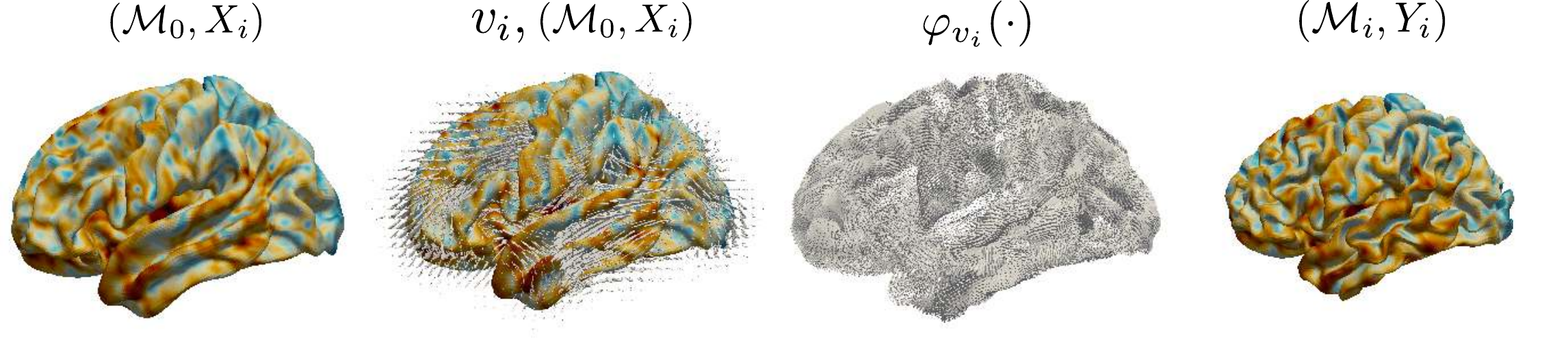}
\caption[]{An illustration of the generation of a \FoS\ through Model~\ref{eq:generation} with $\varphi$ the diffeomorphic deformation operator. From left to right, in the first panel we have a functional sample $X_i$ on the geometric template $\Mtemp$. In the second panel we have a vector field $v_i \in \calV$, a sample of the random function $V$, evaluated on a uniform grid in $\R^3$. This is shown together with $(\Mtemp, X_i)$. In the third panel we have the diffeomorphic deformation $\varphi_{v_i}$, obtained from $v_i$ as described in Section~\ref{sec:deformation_operator}, here displayed as the set of vectors $\{\varphi_{v_i}(\xi_k)\} \subset \R^3$ with $\{\xi_k\}$ the nodes of the triangulated surface representing the template $\Mtemp$. In the fourth panel, we have the \FoS\ $(\M_i, Y_i)$ obtained by applying the deformation $\varphi_{v_i}$ to $\Mtemp$ and `transporting' the functional values with it.}
\label{fig:generation}
\end{figure}

We now describe the \FoSs\ generation process from Model~\ref{eq:generation}, for different choices of the deformation operator:
\begin{itemize}
\item \textit{Shift operator}: Let $\calV = \R^3$, we define $\varphi_v$ to be such that $\varphi_v(x) = x + v$ for all $v \in \calV, x \in \R^3$. Clearly, in this case, $\{\M_i = \varphi_{v_i} \circ \Mtemp\}$ in Model~\ref{eq:generation} would generate a collection of surfaces shifted in the directions specified by $\{v_i\}$.

\item \textit{Identity operator}: Let $\calV$ be the space of smooth functions $v: \Mtemp \ra \R^3$ and let $\varphi_v(x) = x + v(x)$ for all $x \in \Mtemp$. In this case, $\{\M_i\}$ would be a collection of smoothly deformed versions of the template $\Mtemp$. Note however, that the maps being only smooth and not homeomorphic, it cannot be guaranteed that every choice of $v \in \calV$ preserves the topology of $\Mtemp$. Nevertheless, this choice might still represent a valid option in a small deformations setting.

To solve this problem, we could think of restricting $\calV$ to contain only smooth and homeomorphic functions, however, in this way, the linearity of the space $\calV$ is lost, and this is a property of fundamental importance to the subsequent analysis, given that we want to apply linear statistics on the random function $V$, which takes values on $\calV$.

\item \textit{Diffeomorphic operator}: Let $\calV$ be a Sobolev space of sufficiently smooth vector fields from $\R^3$ to $\R^3$ vanishing, with their derivatives, at infinity. Let $\varphi$ be a diffeomorphic deformation operator, i.e. an operator such that $\varphi_v$ is a diffeomorphism of $\R^3$ for all $v \in \calV$. Then, for different choices of $v$, Model~\ref{eq:generation} would generate a collection of surfaces that are diffeomorphic (and thus homeomorphic) deformations of the template $\Mtemp$. More importantly, these deformations are parametrized by the linear space $\calV$, where linear statistics can be applied. For this choice, an illustration of the generative process is shown in Figure~\ref{fig:generation}. The diffeomorphic deformation operator can be defined by means of an Ordinary Differential Equation (ODE). Details of this are described in Section~\ref{sec:deformation_operator}.
\end{itemize}

More complicated generative models could be built from Model~\ref{eq:generation}. For example, the functions $\{v_i\}$ and $\{X_i\}$, representing respectively geometries and functions, could be modelled in terms of conditional expectation of different sources of information on the subjects such as age, status disease  or other subject-specific explanatory variables, as done, in the case of functional data located on 1D domains, in \cite{Hadjipantelis2015}. However, Model~\ref{eq:generation} is the simplest model enabling a comprehensive study of the relation between geometric and functional variability.

\subsection{Geometric and Functional variability}\label{sec:geo_fun_variability}
Here we formalize the geometric and functional variability relationship. Recalling the definition of geometric and functional variability, given in the introduction,  we can notice that in Model~\ref{eq:generation} we have that $\{v_i\}$ describe the geometric variability in the data, while $\{X_i\}$ describe the functional variability in the data. The key idea of this work is to formalize geometric and functional variability by means of functional Principal Component Analysis (fPCA), so that geometric variability can be represented in terms of the Principal Components (PCs) of the random function $V$, generating the samples $\{v_i\}$, and functional variability can be represented in terms of the PCs of the random function $X = \mu + \delta Z$ generating the random samples $\{X_i\}$. Under the hypothesis that a finite number of PCs is sufficient to represent $V$ and $X$, we can than use classical multivariate statistics, such as multivariate regression or canonical correlation analysis, to model the relation between the PCs of $V$ and the PCs of $X$, ultimately formalizing the concept of geometric and functional variability being related. This should also further clarify the choice to introduce a deformation operator $\varphi$. In fact, as already mentioned, the deformation operator allows us to parametrize the space of deformations through the linear space $\calV$, and thus linear fPCA can be applied on the $\calV$-valued random variable $V$.

More formally, under typical assumptions on $V$, thanks to fPCA, $V$ can be expanded in terms of the orthonormal sequence of eigenfunctions $\{\psi^G_j\}$ of the covariance operator of $V$, as
\[
V = \sum_{j=1}^\infty a^G_j \psi^G_j,
\]
where $a^G_1,a^G_2,\ldots$ are uncorrelated real random variables, with variances in decreasing order $\kappa^G_1, \kappa^G_2,\ldots$.  The collection $\{\psi^G_j\}$ defines the strongest modes of variation of the random function $V$ and these are called PC functions. We refer to $\psi^G_j$ as the $j$th mode of geometric variation, or alternatively the $j$th geometric PC function. This represents variations of the type $c \psi^G_j$ around the mean of $V$, with $c \in \R$. The PC function $\psi^G_j$ is thus associated to the geometric deformations $\varphi_{c \psi^G_j}$ of $\R^3$, that applied to the geometric template correspond to the geometries described by $\varphi_{c \psi^G_j} \circ \Mtemp$. In practice, we visualize the $j$th mode of geometric variation by visualizing the associated geometries for some specific choice of $c$, e.g. $\varphi_{\pm \sqrt{\kappa^G_j} \psi^G_j} \circ \Mtemp$. An example of this visualization is given in Figure~\ref{fig:HCP_tPCA}. PCA has been previously used in a similar fashion in \cite{Vaillant2004} and \cite{Tward2017}, with $\varphi$ the diffeomorphic deformation operator, to represent anatomical geometries.

With analogous considerations, the random variable $X$ can be expanded, using the associated orthonormal eigenfunctions $\{\psi^F_j\}$, of the covariance operator of $X$, as
\[
X = \mu + \sum_{j=1}^\infty a^F_j \psi^F_j,
\]
where the real random variables $a^F_1,a^F_2,\ldots$ are uncorrelated with variances, in decreasing order, $\kappa^F_1, \kappa^F_2,\ldots$. We refer to $\psi^F_j$ as the $j$th mode of functional variation.

FPCA basis expansions have the fundamental property of separating the discrete set of stochastic terms from the functional terms. Hence, the relation between the geometry and the functional terms can be formalized in terms of the random variables $\{a^G_j\}$ and $\{a^F_j\}$. We assume that only a finite number of the PC functions are necessary to describe the phenomenon in hand and denote with $\vect{a}^G$ the associated $K^G$-dimensional random vector $(a_1^G, \ldots, a_{K^G}^G)$ and with $\vect{a}^F$ the $K^F$-dimensional random vector $(a_1^F, \ldots, a_{K^F}^F)$.

Different multivariate statistical models can be applied at this stage, to formalize the geometric and functional variability relation in terms of the relation between the random vectors $(a_1^G, \ldots, a_{K^G}^G)$ and $(a_1^F, \ldots, a_{K^F}^F)$.
A first possible formalization of the geometric and functional variability relation is
\begin{equation}\label{model:joint}
\E[X | V] = \mu + \sum_{j=1}^{K^F} \E [a_j^F| \vect{a}^G] \psi_j^F.
\end{equation}
Under linear assumptions on the dependency, the conditional expectation term can be modelled as
\[
\E [a_j^F| \vect{a}^G] = \boldsymbol{\beta}'_j \vect{a}^G,
\]
with $\boldsymbol{\beta}_j$ the $K^G$-dimension vector of the regression coefficients of the $j$th functional mode of variation.

The model above describes how the main modes of geometric variation explain each mode of functional variation, implying that we expect the geometry to influence the functions. This might be the case of neurodegenerative disease, where we expect the functional activity (the function) to adapt to the disease progression (the geometry). However, the reverse roles of geometry on functions is also plausible in some cases. For instance, through a comparative study between taxi drivers and bus drivers, it has been shown that the different functional activation patterns influence the growth of the gray matter volume, and thus the brain geometry \citep{Maguire2006}. Moreover, given that in model (\ref{model:joint}) each mode of functional variations is explained thorugh a linear combination of the modes of geometric variability, the interpretability of the overall model strongly relies on the interpretability of the singular functional main modes of variations.

A second possible formalization of the geometric and functional variability relationship might consist of simply examining the maximal directions of correlation between geometry and function. This is equivalent to performing a Canonical Correlation Analysis (CCA). A CCA analysis on the coefficients of the fPCA basis expansion is equivalent to finding a new basis expansions for $V$ and $X$ as a linear combination of the respective fPCA basis. However, the elements of the new basis are ordered in a way that maximizes the correlation between their coefficients, i.e. the interdependency between geometry and function, representing how the geometric variability associates with the functional variability and vice versa.

\subsection{The diffeomorphic deformation operator}\label{sec:deformation_operator}
The deformation operator, introduced in the Section~\ref{sec:methodology_def}, has to be chosen in such a way that it is flexible enough to represent the observed surfaces, as a deformation of the template surface. Clearly, the shift operator is not sufficient to capture the variations in geometry of the \FoSs\ in Figure~\ref{fig:cortical_surfaces}, in terms of deformation of the template. However, this operator should only include `sensible' deformations, in the sense that the deformation operator should have its image contained in  the set of diffeomorphic deformations from $\R^3$ to $\R^3$. This choice is driven by the fact that diffeomorphic deformations are smooth deformations that preserve the topological properties of the shapes and that avoid two separate points on the template collapsing to one point on the observed surface.

For this reason, we rely on the idea of constructing diffeomorphic deformations as flows of an ODE \citep{Dupuis1998}, which can be parameterized by a Hilbert function space. Specifically, let now $\calV$ be a Sobolev space of sufficiently smooth vector fields from $\R^3$ to $\R^3$ vanishing, with their derivatives, at infinity. Let $v: [0,1] \times \R^3 \ra \R^3$ be a \textit{time dependent} vector field in $L^2([0,1],\calV)$, the space of vector fields with finite (squared) norm $\int_0^1 \| v_t \|^2_\calV dt$.  Then, for a given $v$, the solution $\phi_v:[0,1] \times \R^3 \ra \R^3$ of the ODE
\begin{equation}\label{eq:ODE}
\frac{\partial \phi_v}{\partial t}(t,x) = v_t \circ \phi_v(t,x)\ \qquad t \in [0,1], x \in \R^3.
\end{equation}
with initial conditions $\phi_v(0,x) = x$, is a smooth diffeomorphic map in $\Diff(\R^3)$, at each fixed time $t$ \citep[see, e.g.,][]{Younes2010}. The ODE (\ref{eq:ODE}) is intuitively defining the solution $\phi_v$ to be a function such that, for all $t \in [0,1]$, the time-derivative $\frac{\partial \phi_v}{\partial t}(t,x)$ (i.e. the velocity field at time $t$) is given by the vector field $v_t \circ \phi_v(t,x)$. In other terms $\phi_v$ represents the `flow' described by the velocity vector field $\{v_t:t \in [0,1]\}$.

Note that we use  $\varphi$ and $\phi$ to represent two different object and their relation is defined as follows.

\begin{figure}[!htb]
\centering
\includegraphics[width=1\textwidth]{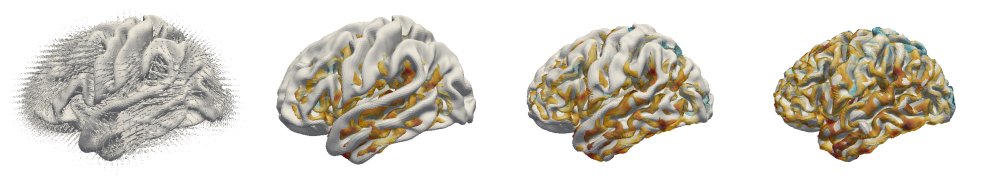}
\caption[]{From left to right, in the first panel we have an initial vector field $v_0 \in \calV$ and in gray the template $\Mtemp$. In the consecutive panels, we show the solution $\phi_v$ of the ODE at the times $t = 0, 0.5, 1$ (which are diffeomorphic deformations of $\R^3$), as deformations of the template $\Mtemp$.  In this specific case, the initial vector field $v_0$ has been chosen in such a way that the surface $\varphi_{v_0} \circ \Mtemp$ is a close approximation of a target surface, i.e. the colored surface in the figure.}
\label{fig:diff_op}
\end{figure}

Given an initial vector field $v_0$, we define $\{v_t:t \in [0,1]\}$ to be the time-variant vector field  which minimizes the quantity $\int_0^t \|v_t\|^2_\calV dt$. For this choice, the vector field $\{v_t:t \in [0,1]\}$ can be derived from $v_0$ through the resolution of the EPDiff equation \citep{Miller2006}. Finally, the deformation operator can be defined to be $\varphi_{v_0}(x) = \phi_v(1,x)$, where $v_0 \in \calV$ is the initial vector field generating $\{v_t:t \in [0,1]\}$, through the EPDiff equation, and $\phi_v$ is the solution of the ODE (\ref{eq:ODE}). The choice to define $\varphi_{v_0}(x)$ with the solution of the ODE (\ref{eq:ODE}) at time $t=1$ is arbitrary, in fact any other choice of a fixed $t > 0$ would have been equivalent, given that the $\phi_v(t,x)$ is guaranteed to be diffeomorphism of $\R^3$ for any $t > 0$.

A summary of the main elements necessary to define $\varphi$ is given by the following
\begin{equation}\label{eq:gen_diff_steps}
\underbrace{v_0 \xrightarrow[\text{EPDiff}]{} \{v_t:t \in [0,1]\} \xrightarrow[\text{ODE (\ref{eq:ODE})}]{} \phi_v \xrightarrow[]{} \varphi_{v_0}:=\phi_v(1,\cdot)}_{\varphi_{v_0}: \R^3 \ra \R^3}.
\end{equation}
In Figure~\ref{fig:diff_op} we show the solution of the ODE (\ref{eq:ODE}) for a given initial vector field $v_0$.
We emphasize that the ODE (\ref{eq:ODE}) is not used here to model the phenomenon in hand, but it is just a convenient tool to generate a  diffeomorphism of $\R^3$ from a smooth vector field $v_0$ belonging to the linear space $\calV$.

\section{Estimation framework}\label{sec:estimation_framework}
The arguments made in the previous section are formalized in terms of quantities derived from the underlying unknown random variables modelling the data generation. However, in practice, only a set of observed noisy \FoSs\ is available, and those quantities have to be estimated from the data. In this section, we mostly work with the set of idealized \FoSs\ $\{(\M_i, Y_i): i=1,\ldots,n\}$. Instead, when the specific computer representation is of importance to the proposed algorithms, we work with the associated collection of pairs denoted with $\{(\M_i^\calT, Y_i^\calT): i=1,\ldots,n\}$, each composed by a triangulated surface $\M_i^\calT \subset \R^3$, approximating the underlying smooth two-dimensional manifold $\M_i \subset \R^3$, and a real piecewise linear function $Y_i^\calT \in L^2(\M_i^\calT)$ representing a noisy approximation of the underlying smooth function $Y_i \in L^2(\M_i)$.

\begin{figure}[!htb]
\centering
\includegraphics[width=1\textwidth]{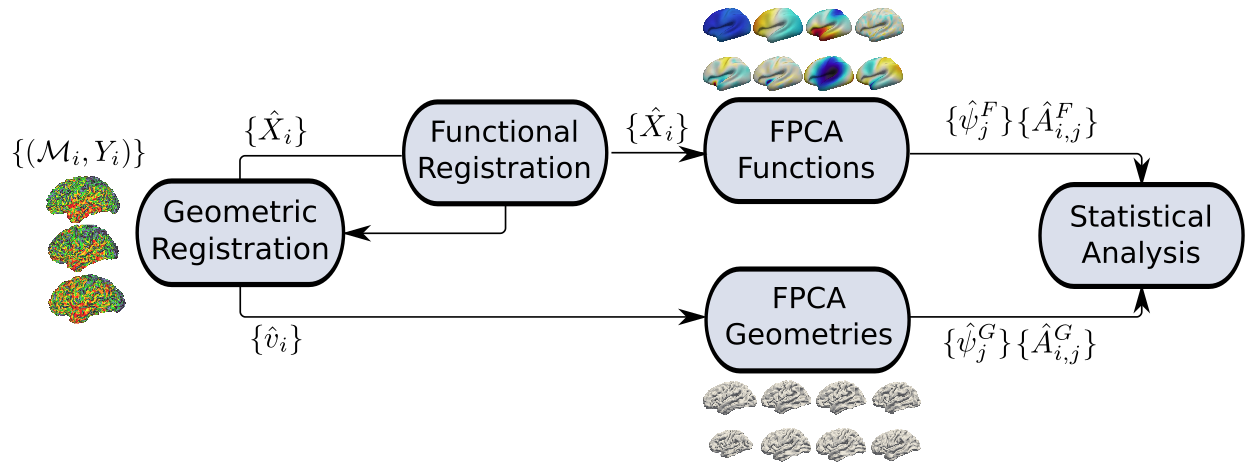}
\caption[]{A flow chart summarizing the main steps of the estimation procedure proposed in Section~\ref{sec:estimation_framework}. In the Geometric Registration step, as described in Section~\ref{sec:geometric_registration_brief}, the FoSs are registered to a template $\M_0$, i.e. for each FoS we estimate a vector field $\hat{v}_i$ representing the geometric object $\M_i$ as the deformed template $\varphi_{\hat{v}_i} \circ \Mtemp$. Each function $\hat{X}_i$ is obtained by transporting $Y_i$ on the template through the estimated registration. In the Functional Registration step, as described in Section~\ref{sec:functional_registration_brief}, the functional information is used to achieve a more accurate registration leading to a corrected version of the functions $\{\hat{X}_i\}$ and the vector fields $\{\hat{v}_i\}$. In the FPCA Functions and FPCA Geometries steps, as described in Section~\ref{sec:fPCA_brief}, fPCA is performed on the functions $\{\hat{X}_i\}$ and the vector fields $\{\hat{v}_i\}$ respectively, leading to the estimation of a set of PC functions and scores representing functional variability and geometric variability. Finally, as described in Section~\ref{sec:geo-fun_emp}, classical statistical analysis is performed on the PC scores $\hat{A}_{i,j}^F$ and $\hat{A}_{i,j}^G$ to study the relation between functional and geometric variability.}
\label{fig:flowchart}
\end{figure}

In this section, we outline the estimation procedures applied to the data to recover the different quantities in Model~\ref{eq:generation}. A flow chart summarizing the main steps is shown in Figure~\ref{fig:flowchart}. The implementation details are covered in the appendices.

\subsection{Geometric Registration and Linear representation of shapes}\label{sec:geometric_registration_brief}
In practice the computation of a diffeomorphic deformation between the template $\Mtemp$ and the surface $\M_i$ is achieved by solving a minimization problem of the form
\begin{equation}\label{eq:minimization_geo}
\hat{v}_i = \argmin_{v_i \in \calV} D^2(\varphi_{v_i} \circ \calM_0, \calM_i) + \lambda \|v_i\|^2_\calV,
\end{equation}
where $D^2(\varphi_{v_i} \circ \calM_0, \calM_i)$, the shape similarity function, is a measure of the amount of mismatching between the deformed template surface and the target surface. The constant $\lambda$ is a weighting parameter between the data-fidelity term and the term $\|v_i\|^2_\calV$, which could be regarded as a measure of the amount of deformation induced by $\varphi_{v_i}$. The functions $\{\hat{v}_i\}$ are an estimation of $\{v_i\}$ in Model~\ref{eq:generation}. In Figure~\ref{fig:diff_op} we show an example of a vector field in $\calV$, estimated by solving (\ref{eq:minimization_geo}), with the aim of representing a target surface as a deformation of a template.

The procedure described is also referred to as the registration step, as in fact the estimated map $\varphi_{\hat{v}_i}$, up to approximation error, defines a one-to-one smooth correspondence between the points of the target manifold $\calM_i$ and the template $\Mtemp$. Thus, the function $\hat{X}_i \in L^2(\Mtemp)$, obtained by registering $Y_i$ to the template, can be defined as the element $\hat{X}_i$ such that $\hat{X}_i(x) = Y_i(\varphi_{\hat{v}_i}(x))$ for all $x \in \Mtemp$. The registered maps $\{\hat{X}_i\}$ can be regarded as a first approximation of $\{X_i\}$ in Model~\ref{eq:generation}. In practice, there might be a small approximation error between $\varphi_{v_i} \circ \calM_0$ and $\calM_i$, which might contrast with the definition $\hat{X}_i(x) = Y_i(\varphi_{\hat{v}_i}(x))$ for all $x \in \Mtemp$, as $\varphi_{\hat{v}_i}(x)$ might not exactly belong to $\M_i$. However, we assume that $\varphi_{\hat{v}_i}(x)$ is close enough to $\M_i$, for all $x \in \Mtemp$, and in practice define $\hat{X}_i(x) = Y_i(y)$ with $y \in \M_i$ the nearest neighbor of $\varphi_{\hat{v}_i}(x)$.

The implementation of the registration algorithm (\ref{eq:minimization_geo}) requires the definition of a shape similarity function $D$. As already mentioned, the geometry of a \FoS\ is in practice encoded as a triangulated mesh, we thus define the similarity function $D$ between triangulated surfaces. We should differentiate between two possible settings at this point. In the first setting, we suppose that a correspondence between the points of the triangulated surfaces $\calM_0^\calT$ and $\calM_i^\calT$ is known for all $i=1,\ldots,n$. In other terms, we suppose that $\calM_0^\calT$ and $\calM_i^\calT$ have already been registered and thus there is a set of landmarks $\{x_l, y_l: x_l \in \calM_0^\calT, y_l \in \M_i^\calT\}$  in correspondence between them. In this case, a simple mismatching functional is given by the Euclidean distance between the correspondent landmarks i.e.
\begin{equation}\label{eq:landmarks}
D^2(\varphi_{v_i} \circ \calM^\calT_0, \calM^\calT_i) = \sum_l \| \varphi_{v_i}(x_l) - y_l \|^2_{\mathbb{R}^3}.
\end{equation}
This choice has been adopted for instance in \cite{Joshi2000}.

This situation is frequent in neuroimaging, a field that has developed their own ad hoc registration algorithms and where diffeomorphic constraints are explicitly imposed without the necessity to use a diffeomorphic deformation operator. In this case the estimates $\{\hat{X}_i\}$ are already provided, given that $\calM_0^\calT$ and $\calM_i^\calT$ have already been registered, nevertheless, the framework introduced here is still of relevance, in fact, we still need to estimate $\{\hat{v}_i\} \subset \calV$, in equation (\ref{eq:minimization_geo}), to represent the given registration maps (i.e. deformation maps), and thus the geometries, in terms of elements of a linear space, which is a fundamental property to the subsequent analysis.

In the second setting, we suppose that a registration step has not been performed yet. In this situation, registration and linear representation can be performed jointly by choosing an appropriate shape similarity function $D$ not based on landmarks, but for instance, proximity. An example of such similarity function is proposed in \cite{Vaillant2005} and \cite{Vaillant2007}, and is defined as follows. Let $K_{\calZ}: \R^3 \times \R^3 \ra \R^{3 \times 3}$ be a Gaussian isotropic kernel of variance $\sigma^2_\calZ$, i.e. $K_{\calZ}(x,y)= \exp(-\| x-y \|_2^2/ (2\sigma_{\calZ}^2)) \Id_{3 \times 3}$, with $\Id_{3 \times 3}$ denoting a $3 \times 3$ identity matrix. Indeed, such kernel can be any symmetric positive definite kernel, however it is common to choose a Gaussian kernel.
Denote with $c(l)$ and $\eta(l)$, respectively, the center point and the normal vector of the $l$th triangle of the mesh $\varphi_{v_i} \circ \M_0^\calT$.  Denote with $c_i(q)$ and $\eta_i(q)$, respectively, the center point and the normal vector of the $q$th triangle of the mesh $\calM^\calT_i$. Moreover, let the triangles of the mesh $\varphi_{v_i} \circ \M_0^\calT$ be indexed by $l$ and $g$ and the triangles in $\calM^\calT_i$ be indexed by $q$ and $r$. The resulting shape similarity function has the form
\begin{align}\label{eq:empirical}
\begin{split}
D^2(\varphi_{v_i} \circ \calM^\calT_0, \calM^\calT_i)& = \sum_l \sum_g K_{\calZ} (c(l),c(g)) \eta(l) \cdot \eta(g)\\
&-2 \sum_l \sum_q K_{\calZ}(c(l),c_i(q)) \eta(l) \cdot \eta_i(q)\\
&+ \sum_q \sum_r K_{\calZ}(c_i(q),c_i(r)) \eta_i(q) \cdot \eta_i(r),
\end{split}
\end{align}
with $\cdot$ denoting the scalar product in $\R^3$.
Intuitively, the first and last terms measure deformations to the local geometry within the two surfaces, and the middle term measures the mismatch in local
geometry between the two surfaces.

Thanks to the procedure outlined in this section, given a set of \FoSs, we are able to register them to a fixed template $\Mtemp$. As a result, the information regarding the geometry of the data is stored in terms of the estimates $\{\hat{v}_i\} \subset \calV$ of $\{v_i\}$ in Model~\ref{eq:generation}. These are estimated so that $\varphi_{\hat{v}_i} \circ \Mtemp$ resembles the geometry of the $i$th \FoS. Moreover, we obtain a set of functions $\{\hat{X}_i\}$ on the fixed template, that are a first estimate of the functions $\{X_i\}$ in Model~\ref{eq:generation}.

In practice the space of smooth functions $\calV$ is implemented as a Reproducing Kernel Hilbert Space (RKHS), as described in Appendix~\ref{app:diffeomorphic_registration}.

\subsection{Functional Registration}\label{sec:functional_registration_brief}
The aim of this section is the introduction of a novel functional registration algorithm for functional data whose domain is a fixed two-dimensional manifold. The functional registration algorithm can then be applied to align the set of functions $\{\hat{X}_i: \hat{X}_i \in L^2(\Mtemp)\}$, estimated in Section~\ref{sec:geometric_registration_brief}, by registering them to a template function $X_0 \in L^2(\Mtemp)$, which can be in first instance approximated by the cross-sectional sample mean of $\{\hat{X}_i\}$. The rationale for such a procedure is that, as well known in FDA, the functions $\{\hat{X}_i\}$ on $\Mtemp$ should in principle be able to drive a better registration, on the assumption that the underlying functions $\{X_i\}$ in Model~\ref{eq:generation} have a preponderant mean effect, with respect to its second order variation.

In fact, each estimated function $\hat{X}_i$, strongly depends on the associated deformation map $\varphi_{\hat{v}_i}$, whose estimation is usually driven only by geometric features. Hence, a systematic mis-registration, due to a naive approximation of the deformation maps, could introduce fictitious functional variability on the functions $\{\hat{X}_i\}$, which in fact should be accounted for by geometric variability, in particular in a setting where obvious landmarks are not available and the deformations $\{\varphi_{\hat{v}_i}\}$ are estimated while ignoring the functional information. The functional registration algorithm can be regarded as a correction step to $\{\hat{X}_i\}$, and thus $\{\hat{v}_i\}$, estimated from Section~\ref{sec:geometric_registration_brief}.

A review on the registration of functional data can be found in \cite{marron2015}. However, most of the FDA literature treats only the case of functions whose domain is an interval of the real line. Registration of 2D images has also been well studied \citep[see e.g.,][for a review]{Zitova2003}. Methods that preserve invertibility of the deformation have also been proposed for 2D/3D Euclidean images \citep{Vercauteren2009} and extended to functions with spherical domains in \cite{Yeo2010}. However, to the best of our knowledge, these methods are not able to deal with the registration of a collection of functions whose domain is a fixed generic two-dimensional manifold embedded in $\R^3$.

Alternatively, in the case of landmark based registration, functional information can be introduced into the registration process, by modifying the algorithm that provides the landmarks, to account for function similarity. In the case where landmarks are not available functional information can be introduced by equipping the shape similarity functional (\ref{eq:empirical}) with a functional similarity term, as done in \cite{Charon2014} and \cite{Charlier2017}.

\subsubsection{Definitions}\label{sec:functional_registration_def}
\begin{figure}[!htb]
\centering
\includegraphics[width=0.4\textwidth]{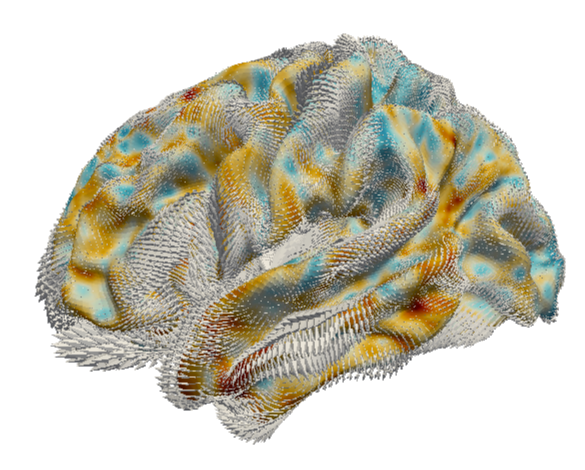}
\caption[]{A section of the tangent bundle $T \Mtemp$, which has been computed by minimizing the linearized version of the equation (\ref{eq:diff-nonlinear}).}
\label{fig:fun_vf}
\end{figure}

Let $T_p \Mtemp$ be the tangent space on the point $p \in \Mtemp$ and let $g_p$ be the metric on $\Mtemp$, i.e. a scalar product on the tangent space $T_p \Mtemp$. In our case it is natural to consider the scalar product induced by the Euclidean embedding space $\R^3$, i.e. the first fundamental form. Define the tangent bundle to be the disjoint union of tangent spaces $T \Mtemp = \dot\bigcup_{p \in \Mtemp} T_p \Mtemp = \bigcup_{p \in \Mtemp} \{p\} \times T_p \Mtemp$. A section of the tangent bundle $T \Mtemp$ is the formalization of the concept of a vector field on $\Mtemp$, an example of which is shown in Figure~\ref{fig:fun_vf}. We denote with $L^2(T\Mtemp)$ the Hilbert space of square integrable sections of $T\Mtemp$. Moreover, let $\Delta_{BL}$ be the Bochner-Laplacian operator. The Bochner-Laplacian of a smooth vector field $v$, i.e. $\Delta_{BL} v$, is a vector field on $\Mtemp$, whose $L^2$ norm gives a measure of the smoothness of the vector field $v$. i.e. low values for smooth vector fields $v$ and high values for rough vector fields. A more formal definition of the Bochner-Laplacian operator, from the Levi-Civita operator, is given in Appendix~\ref{app:bochner-laplacian}.

\subsubsection{Estimation}\label{sec:functional_registration_est}
The registration of $\{\hat{X}_i\}$ is performed in an iterative fashion, which means that each function $\hat{X}_i$ is aligned to the function $X_0$ by composition of small diffeomorphic deformations. Let $\{s_i: \Mtemp \ra \Mtemp\}$ be the set deformation maps estimated from the previous iterations of the algorithm, such that $\hat{X}_i \circ s_i$ is a registered version of $\hat{X}_i$ to the function $X_0$. The functions $\{s_i\}$ can be the set of identity maps in the first iteration. Moreover, let $\{p_j: j=1,\ldots,S\} \subset \Mtemp$ be a collection of $S$ control points where the functions $\{\hat{X}_i\}$ are sub-sampled. In practice, these will be the nodes of the triangulation $\calM_0^\calT$, i.e. the points where the functions are actually observed.

With a slight abuse of notation, let the diffeomorphic function $\phi_{u}: \Mtemp \ra \Mtemp$ be the solution generated at time $t=1$ by the ODE
\begin{align}\label{eq:ODE-stationary}
\begin{cases}
\frac{\partial \phi_{u}}{\partial t}(t,x) = u \circ \phi_{u}(t,x)\ \qquad &t \in [0,1], x \in \Mtemp,\\
\phi_{u}(0,x) = x \qquad &x \in \Mtemp
\end{cases}
\end{align}
where $u$ is a sufficiently smooth vector field on $\Mtemp$. If $\Mtemp$ has a boundary, than we assume $u$ vanishes, with their derivatives, on the boundary. Such an ODE is used here as a tool to generate a diffeomorphic function $\phi_{u}$ from a vector field $u$ that needs only to be smooth. Then, we propose to estimate a set of functional registration maps, each aligning $\hat{X}_i \circ s_i$ to $X_0$ by minimizing
\begin{equation}\label{eq:diff-nonlinear}
E_{\Mtemp}(u_i) = \sum_{j=1}^S \big (X_0(p_j) - \hat{X}_i \circ s_i \circ \phi_{u_i} (p_j) \big)^2 +  \lambda\|\Delta_{BL} u_i\|^2_{L^2(T\Mtemp)},
\end{equation}
where $\|\Delta_{BL} u_i\|^2_{L^2(T\Mtemp)}$ is the $L^2$ norm of the vector field $\Delta_{BL} u_i$, which imposes smoothness on $u_i$. The constant $\lambda$ is a weighting coefficient between the data fidelity term, i.e. how well aligned we want $\hat{X}_i \circ s_i \circ \phi_{u_i}$ to be to $X_0$, and the smoothing term, i.e. how smooth we want the vector field $u_i$ to be.

The term $\hat{X}_i \circ s_i \circ \phi_{u_i}$ in Equation (\ref{eq:diff-nonlinear}), is then linearized with respect to $u_i$. This results in the approximation
\[
\hat{X}_i \circ s_i \circ \phi_{u_i} \approx \hat{X}_i \circ s_i + L_{u_i},
\]
where $L_{u_i}$ is a first order approximation of $\hat{X}_i \circ s_i \circ \phi_{u_i} - \hat{X}_i \circ s_i$. By means of Vector Finite Elements, an approximate solution $\hat{u}_i$, at the nodes of $\M^\calT_0$, can be characterized in terms of the solution of a linear system. An approximate vector field $\hat{u}_i$ on the triangulation $\M^\calT_0$ is then computed by linear interpolation of the solution found at the nodes of the triangulation. Details of this procedure can be found in Appendix~\ref{app:functional_reg}. The main steps of the functional registration algorithm are summarized in Algorithm~\ref{alg:fun-reg}.

\begin{algorithm}[!htb]
\caption{Functional Registration Algorithm}\label{alg:fun-reg}
\begin{algorithmic}[1]
\State Initialization:
\begin{enumerate}[label=(\alph*)]
\item Initialize $\{s_{i}^0 \gets Id: i =1,\ldots,n\}$ to be the identity functions on $\Mtemp$
\item Initialize $\{\hat{X}_i: i=1,\ldots,n\}$ to be the functions estimated from Section~\ref{sec:geometric_registration_brief}
\item Initialize the functional template to be $X_0 \gets \frac{1}{n}\sum_i \hat{X}_i$
\end{enumerate}

\State Compute $\{\hat{u}^{k}_i: i=1,\ldots,n\}$, the solution at the $k$th iteration, from the second order functional

\[
E_{\Mtemp}(u_i) = \sum_{j=1}^S \big (X_0(p_j) - \hat{X}_i \circ s_i^{k-1}(p_j) - L_{u_i}(p_j) \big)^2 +  \lambda\|\Delta_{BL} u_i\|^2_{L^2(T\Mtemp)},
\]

\State Compute the registration maps $\{\phi_{\hat{u}^{k}_i}: i=1,\ldots,n\}$ by solving the ODE
\begin{align*}
\begin{cases}
\frac{\partial \phi_{u_i^k}}{\partial t}(t,x) = u_i^k \circ \phi_{u_i^k}(t,x)\ \qquad &t \in [0,1], x \in \Mtemp,\\
\phi_{u_i^k}(0,x) = x \qquad &x \in \Mtemp
\end{cases}
\end{align*}

\State Update current registration maps and functional template:

$\{s_i^{k} \gets s_i^{k-1} \circ \phi_{\hat{u}^{k}_i}: i=1,\ldots,n\}$

$X_0 \gets \frac{1}{n}\sum_i \hat{X}_i \circ s_i^k$

\State Output and analysis (e.g. fPCA) of the result of the current iteration:

$\{\hat{X}_i \circ s_i^{k}: i=1,\ldots,n\}$

\State Repeat Steps 2--5 until until a stopping criterion is satisfied
\end{algorithmic}
\end{algorithm}

Each iteration of the functional registration algorithm result in a newly estimated set of functions $\{\hat{X}_i \circ s_i^{k}\}$, representing a re-aligned correction of the maps $\{\hat{X}_i\}$. The composition $\hat{X}_i \circ s_i^{k}$ means that for all $x \in \Mtemp$ the functional value $\hat{X}_i(x)$ is, after $k$ iterations, relocated on the point $(s_i^{k})^{-1}(x) \in \Mtemp$. Thus, the functional registration also has the effect of correcting the overall geometric deformations $\{\varphi_{\hat{v}_i}:\Mtemp \ra \M_i\}$, estimated in Section~\ref{sec:geometric_registration_brief}, to be
\begin{equation}\label{eq:overall_def}
\varphi_{\hat{v}_i} \circ (s_i^{k})^{-1}, \qquad i=1,\ldots,n.
\end{equation}

The geometric registration model in Section~\ref{sec:geometric_registration_brief} and the functional registration model, introduced in this section, are similar in spirit, as they both rely on the idea that given a smooth vector field we can generate a diffeomorphic vector field by means of an ODE. However, they also differ in many aspects. For instance, they differ in the way smoothness is imposed. In the geometric registration model, smoothness is imposed by penalizing through the norm of a RKHS. In the functional registration model, within each iteration, smoothness is imposed by means of a differential operator, while the overall smoothness is controlled by the number of iterations. Moreover, in the geometric registration, the ODE is defined for a time-variant vector field, instead in the functional model the ODE is defined for a stationary vector field. Some of these aspects, including the link between penalizing through the norm of a RKHS and penalizing by means of a differential operator, are discussed in Appendix~\ref{app:functional_reg}.

The functional registration model introduced in this section, as opposed to the geometric registration model in Section~\ref{sec:geometric_registration_brief}, is based on the composition of small deformations, where at the $k$th iteration, $\{\hat{X}_i \circ s_i^{k}: i=1,\ldots,n\}$ represent the re-aligned versions of $\{\hat{X}_i \circ s_i^{k-1}: i=1,\ldots,n\}$. The constant $\lambda$ in the model, controls the change between the functions $\{\hat{X}_i \circ s_i^{k-1}\}$ and those estimated at the next iteration, as in fact large values of $\lambda$ privilege small deformations. This has the advantage that the fPCA analysis can be re-performed on the functions $\{\hat{X}_i \circ s_i^{k}\}$ at each iteration $k$. The output of this analysis can provide useful information for the next step of the functional registration algorithm, as for instance a stopping criterion in a similar fashion to \cite{Kneip2008}.

In summary, we have introduced a method that exploits the functional information to achieve a better registration by updating the functional estimates $\{\hat{X}_i\}$ to be $\{\hat{X}_i \circ s_i^{k_\text{stop}}\}$, and the diffeomorphic geometric deformations $\varphi_{\hat{v}_i}$ to be $\{\varphi_{\hat{v}_i} \circ (s_i^{k_\text{stop}})^{-1}\}$, where $k_\text{stop}$ denotes the iteration where the functional algorithm is stopped. As pointed out at many stages in this work, it is however important to have a representation of the final update of the deformations maps $\{\varphi_{\hat{v}_i} \circ (s_i^{k_\text{stop}})^{-1}\}$ in terms of elements of the linear Hilbert space $\calV$, so that we can perform linear statistics. To this purpose, we can estimate such elements by applying the geometric deformation model in Section~\ref{sec:geometric_registration_brief}, i.e. by solving
\[
\hat{v}_i^{k_\text{stop}} = \argmin_{v_i \in \calV} D^2 (\varphi_{v_i} \circ \calM^\calT_0, \varphi_{\hat{v}_i} \circ (s_i^{k_\text{stop}})^{-1} \circ \calM^\calT_0 ) + \lambda \|v_i\|^2_\calV,
\]
where $D^2$ denotes the landmark distance defined in equation (\ref{eq:landmarks}).

The overall procedure in this section results in a set of corrected estimates $\{\hat{X}_i \circ s_i^{k_\text{stop}}\}$ and $\{\hat{v}_i^{k_\text{stop}}\}$, that exploit functional information, estimating respectively $\{X_i\}$ and $\{v_i\}$ in Model~\ref{eq:generation}. To ease the notation, in the next section, we drop the index on the number of iterations of the functional registration algorithm, denoting with $\{\hat{X}_i\}$ and $\{\hat{v}_i\}$ the corrected estimates of functions and geometric deformations respectively.

\subsubsection{Remarks on computational times}\label{sec:remarks_comp_times}
It is also important to highlight that the idea of alternating between each iteration of the functional registration algorithm and the fPCA analysis on the functions is ultimately enabled by the computational efficiency of the proposed functional registration algorithm and fPCA algorithm on the functions. In the case of the application, in Section~\ref{sec:application}, each \FoS\ is represented by a 32K nodes triangulated surface and the associated 32K functional values on the nodes. In this setting, the computational time of one iteration of the functional registration algorithm, applied between two functions, is in the order of $2$ minutes on
a Intel Core i5-3470 3.20GHz workstation, with 4 GB of RAM. The computational time for a singular PC of the functions is 15 seconds, on the same workstation, with the fPCA implementation proposed in Section~\ref{sec:fPCA_brief}. Instead, the landmark driven geometric registration of the 32K nodes template to a 32K nodes surface, representing a cerebral cortex, takes approximately 3 hours on a cluster's node equipped with a Dell T620 server and a NVIDIA K20 GPU.

\subsection{Functional Principal Component Analysis}\label{sec:fPCA_brief}
In Sections~\ref{sec:geometric_registration_brief}-\ref{sec:functional_registration_brief} we introduced the estimation procedure for the objects $\{\hat{v}_i: i=1,\ldots,n\}$ representing the geometries and $\{\hat{X}_i: i=1,\ldots,n\}$ representing the functions, from a set of $n$ \FoSs. In this section, the aim is to outline the estimation procedure to the empirical PC component functions from the observed objects $\{\hat{v}_i\}$ and $\{\hat{X}_i\}$, in analogy to what proposed in Section~\ref{sec:geo_fun_variability}, in terms of PCs of the underlying random functions $V$ and $X$.

\subsubsection{Geometric variability}
The empirical PC functions are in practice computed from the eigen-decomposition of the empirical covariance operator $\hat{C}_\calV$, defined as
\begin{equation}\label{eq:empirical-cov}
\hat{C}_\calV(v) = \frac{1}{n} \sum_{i=1}^n \langle v, \hat{v}_i-\bar{v} \rangle_{\calV} (\hat{v}_i- \bar{v}), \qquad v \in \calV,
\end{equation}
where $\bar{v} = \frac{1}{n} \sum_{i=1}^n \hat{v}_i$ and $\langle \cdot, \cdot \rangle_{\calV}$ denotes the scalar product in $\calV$. An explicit solution of this eigenvalue problem can be derived by expanding $v$ and $\hat{v}_i$ in (\ref{eq:empirical-cov}) over a basis of $\calV$ or discretizing the problem over a fine grid of $\R^3$. Since the number of observations in this setting is small with respect to the size of the space, an appropriate choice of the basis is given by the collection of the actually observed vector fields $\hat{v}_i$ \citep{Ramsay2005}. Thus, the eigenvalue problem $\hat{C}_\calV(\psi^G_j) = \kappa^G_j \psi^G_j$ can be re-formulated to a discrete eigenvector problem in terms of the basis expansion coefficients, leading to the empirical PC functions estimates $\{\hat{\psi}^G_j\}$ and empirical variance estimates $\{\hat{\kappa}^G_j\}$. The empirical PC scores vectors can be estimated by projecting $\{\hat{v}_i\}$ on the estimated PC functions, i.e. the $i$th element of the $j$th scores vector is given by
\[
\hat{A}^G_{i,j} = \langle \hat{v}_i - \bar{v}, \hat{\psi}^G_j \rangle_\calV \qquad i=1,\ldots,n, \, j=1,\ldots,K^G.
\]

The empirical mean $\bar{v}$ can be neglected, as the underlying random function is assumed to have zero mean. The empirical $j$th mode of geometric variation is thus represented by the PC function $\hat{\psi}^G_j$, which is associated to the deformations $\varphi_{\pm\sqrt{\hat{\kappa}^G_j} \hat{\psi}^G_j}$ of $\R^3$ that applied to the geometric template correspond to the change of geometry described by $\varphi_{\pm \sqrt{\hat{\kappa}^G_j} \hat{\psi}^G_j} \circ \Mtemp$. The observed vector fields can be finally expressed in terms of the basis expansion, also known as the Karhunen-Lo{\`e}ve expansion:
\begin{equation}\label{eq:vector_field_basis}
\hat{v}_i \approx \sum_{j=1}^{K^G} \hat{A}_{i,j}^G \hat{\psi}_j^G.
\end{equation}
Equation (\ref{eq:vector_field_basis}) emphasizes the fact that the matrix $(\hat{A}_{i,j}^G)_{ij}$ is such that the $i$th row is a compact description of the vector field $\hat{v}_i$.

\subsubsection{Functional variability}
From similar arguments, we can build an estimator for the PC functions and PC scores vectors for the functions $\{\hat{X}_i\}$. 
The estimated functions $\{\hat{X}_i\}$ are noisy estimates of the realization of the underlying unobserved random function $X$. A pre-smoothing of the noisy functions could be considered, however here we rely on the fPCA algorithm proposed in \cite{Lila2016}, where the regularization term is applied directly to the PC functions to be estimated.

In fact, the PC functions $\{\psi^F_j\}$ of the centered random function $X - \mu$, satisfy the following property
\begin{equation}\label{eq:FPCA_minimization}
\{\psi^F_m\}_{m=1}^M =
\argmin_{(\{\psi_m\}_{m=1}^M: \langle \psi_m, \psi_l \rangle_{L^2(\Mtemp)} = \delta_{ml})}
\E \int_{\Mtemp}\bigg\{X - \mu - \sum_{m=1}^M \langle X - \mu, \psi_m \rangle_{L^2(\Mtemp)} \psi_m \bigg\}^2,
\end{equation}
where $\int_{\Mtemp}$ denotes the surface integral over $\Mtemp$ and $\langle \cdot, \cdot \rangle_{L^2(\Mtemp)}$ denotes the scalar product in $L^2(\Mtemp)$. In (\ref{eq:FPCA_minimization}) we can see that the PC functions minimize the loss of information caused by the truncation of the series expansion to the first $M$ components. Let $\{p_j: j = 1,\ldots,S\} \subset \Mtemp$ be a collection of $S$ points where the estimated functions $\{ \hat{X}_i \}$ are sub-sampled. In practice, these will be the nodes of the triangulation $\M_0^{\calT}$, i.e. the points where the functions are actually observed. Let $\Delta$ be the Laplace-Beltrami operator \citep[see e.g.][]{Chavel2006}. The Laplace-Beltrami operator of a smooth function $f \in L^2(\Mtemp)$ is a function in  $L^2(\Mtemp)$ that gives a measure of the local curvature of the function $f$.

The first PC function $\hat{\psi}^F_1 \in L^2(\Mtemp)$ and associated first scores vector $(\hat{A}^F_{1,1},\ldots, \hat{A}^F_{n,1})$ are estimated by minimizing the following regularized empirical version of (\ref{eq:FPCA_minimization}):
\begin{equation}\label{eq:FPCA_minimization_emp}
(\hat{\psi}^F_1, \{\hat{A}^F_{i,1}\}_{i=1}^n)  = \argmin_{\psi_1, \{A_{i,1}\}_{i=1}^n } \sum_{i=1}^{n} \sum_{j=1}^{S} \big(\hat{X}_i(p_j) - \bar{X}(p_j) - A_{i,1} \psi_1(p_j)\big)^2 + \lambda \|\Delta \psi_1\|_{L^2(\Mtemp)},
\end{equation}
where $\bar{X}$ denotes the sample mean function of $\{\hat{X}_i\}$ and $\lambda$ is a weighting coefficient between the empirical and regularizing term. The regularization term imposes smoothness on the estimated PC function $\hat{\psi}^F_1$, coherently with the structure of the manifold $\M_0$.
Subsequent PCs can be estimated by reapplying (\ref{eq:FPCA_minimization_emp}) to the residuals. Details of the implementation and an application to functional Magnetic Resonance Imaging can be found in \cite{Lila2016}.

The observed functions $\{\hat{X}_i\}$ can be finally expressed in terms of the basis expansion
\begin{equation}
\hat{X}_i \approx \bar{X} + \sum_{j=1}^{K^F} \hat{A}_{i,j}^F \hat{\psi}_j^F.
\end{equation}
The matrix $(\hat{A}_{i,j}^F)_{ij}$ is such that the $i$th row is a compact description of the function $\hat{X}_i$.

\subsection{Geometric and Functional variability relation}\label{sec:geo-fun_emp}
The matrices $(\hat{A}_{i,j}^G)_{ij}$ and $(\hat{A}_{i,j}^F)_{ij}$, computed in Section~\ref{sec:fPCA_brief}, are such that their $i$th row represents a compact description of the geometry and functions of the $i$th \FoS\ $(\M_i, Y_i)$. Each row of these matrices could also be regarded as the estimated empirical $i$th realization of the random vector $(a_1^G, \ldots, a_{K^G}^G)$ and $(a_1^F, \ldots, a_{K^F}^F)$ defined in Section~\ref{sec:geo_fun_variability}. As outlined in that section, the matrices $(\hat{A}_{i,j}^G)_{ij}$ and $(\hat{A}_{i,j}^F)_{ij}$ can then be used to study the relation between geometric variability and functional variability of the given collection of \FoSs. To this end we can perform, for instance, a linear regression analysis where we try to explain the $j$th mode of functional variability as a linear combination of the ${K^G}$ modes of geometric variation.

Alternatively, we could perform CCA, and look for the $l$th mode of co-variation $(\hat{\vect{w}}^{G,l}, \hat{\vect{w}}^{F,l})$, representing the $l$th maximally correlated linear combination $\hat{\vect{w}}^{G,l} \in \R^{K^G}$, of the ${K^G}$ modes of geometric variation with the linear combination $\hat{\vect{w}}^{F,l} \in \R^{K^F}$ of the ${K^F}$ modes of functional variation. The $l$th mode of co-variation $(\hat{\vect{w}}^{G,l}, \hat{\vect{w}}^{F,l})$ can be visualized as the sequence of \FoSs\
\begin{align}\label{eq:CCA_sim}
\begin{cases}
\begin{split}
\M_{\text{CCA},l} &= \varphi_{c \; \hat{\psi}^G_{\text{CCA},l}} \circ \Mtemp,\\
Y_{\text{CCA},l} &= c \; \hat{\psi}^F_{\text{CCA},l} \circ \varphi^{-1}_{c \;  \hat{\psi}^G_{\text{CCA},l}},
\end{split}
\end{cases}
\end{align}
obtained by varying $c \in \R$ in an interval containing $0$, with $\hat{\psi}^G_{\text{CCA},l} = \sum_{j=1}^{K^G} \hat{w}^{G,l}_j \hat{\psi}^G_j$ and $\hat{\psi}^F_{\text{CCA},l} = \sum_{j=1}^{K^F} \hat{w}^{F,l}_j \hat{\psi}^F_j$, where $\{\hat{\psi}^G_j\}$ and $\{\hat{\psi}^F_j\}$ are the estimated geometric and functional PC component functions, while $\hat{w}^{G,l}_j$ and $\hat{w}^{F,l}_j$ denote the $j$th element of $\hat{\vect{w}}^{G,l}$ and $\hat{\vect{w}}^{F,l}$ respectively. An example of such a visualization is shown in Figures~\ref{fig:HCP_CCA_1}-\ref{fig:HCP_CCA_2}, for the real application.

\subsection{Choice of the hyper-parameters}
In the proposed models, various hyper-parameters have to be chosen. In particular, in the geometric registration step in Section~\ref{sec:geometric_registration_brief}, we have to choose the regularization weighting parameter $\lambda$. The regularization weighting parameter, in our analysis, does not play a large role. In fact, if the surfaces were noisy reconstructions, its choice would have been more delicate. However, in practice, the surfaces are extracted from a regularized segmentation process of 3D images, and thus are smooth. For this reason, the regularization weighting parameter $\lambda$, in the geometric registration, is chosen to be small.

As previously mentioned, $\calV$ is in practice a RKHS. Important to the registration problem is the choice of $\sigma_\calV$, the size of the kernel of the RKHS $\calV$ (see Appendix~\ref{app:diffeomorphic_registration}). In fact a RKHS with a large kernel size $\sigma_\calV$ is able to better capture large deformations (e.g. size differences), while under-fitting local differences. A RKHS with a small kernel size has an opposite behaviour. Following the approach of \cite{Bruveris2012}, we take a sum of two Gaussian kernels, which allows the space $\calV$ to account for both large and small deformations.

The functional registration has also a regularization weighting parameter $\lambda$, which determines how slowly the algorithm approaches an optimal solution. As in \cite{Kneip2008}, after some experimentation, we choose the value $\lambda$ that achieves a smooth variation on functional PC functions, obtained from the functional variability analysis, between each iteration. To determine the number of iterations needed, we examine the eigenvalue plots (scree plots) to determine when stability of these plots has been reached in a analogous manner to \cite{Kneip2008}.  Finally, the regularization weighting parameter of the fPCA algorithm applied to the functions, has been chosen by $K$-fold cross-validation, with $K = 5$, details of which can be found in \cite{Lila2016}.

On a more general note, choosing the hyper-parameters of the registration algorithms, in a data-driven fashion, is admittedly a very difficult problem and it has been very little explored in the current literature, even in simpler situations such as for functions on the real line. The above only represents one possible method of choosing them, which appears to work well in our application, although further work would be needed for very different settings.

\section{Analysis of a synthetic dataset}\label{sec:simulations}
In this section, we validate the estimation framework introduced in Section~\ref*{sec:estimation_framework}, by performing a study on a dataset generated from Model~\ref{eq:generation}. We thus proceed with defining the unknown quantities of such model. We will not use different notation for the theoretical objects and their respective computer representations, unless necessary.

\begin{figure}[ht]
\centering
\includegraphics[width=0.35\textwidth]{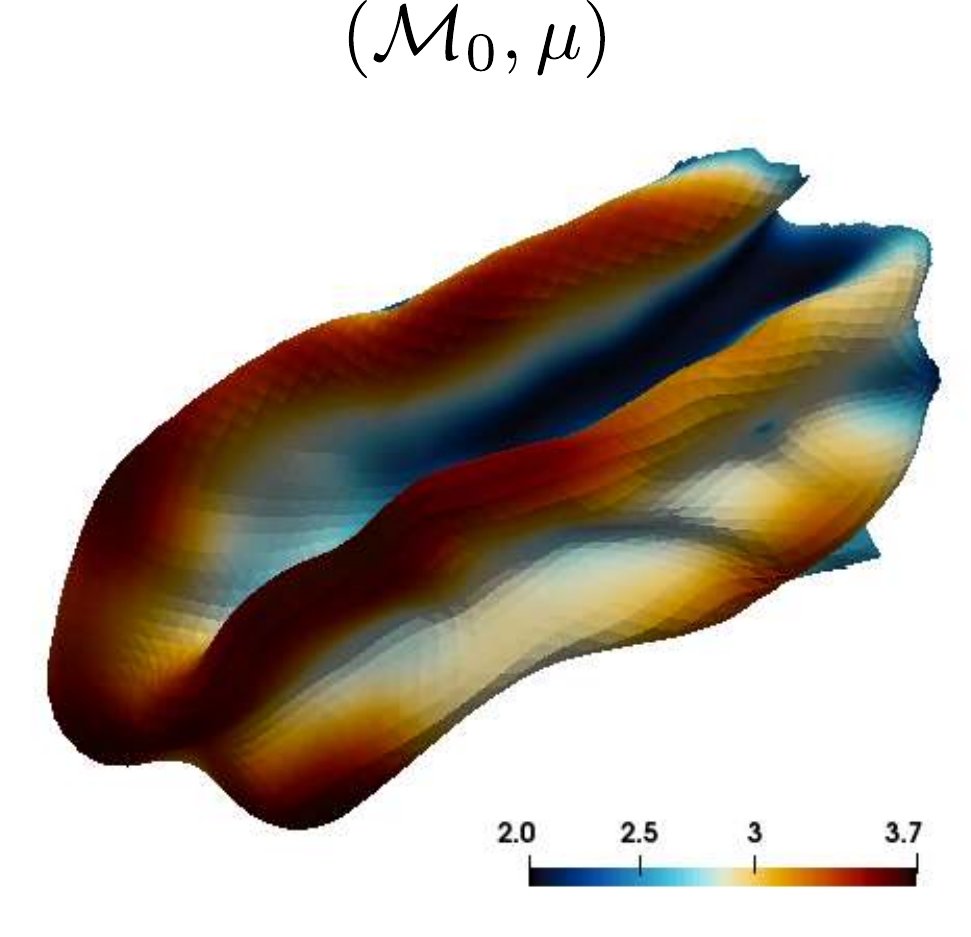}
\includegraphics[width=0.35\textwidth]{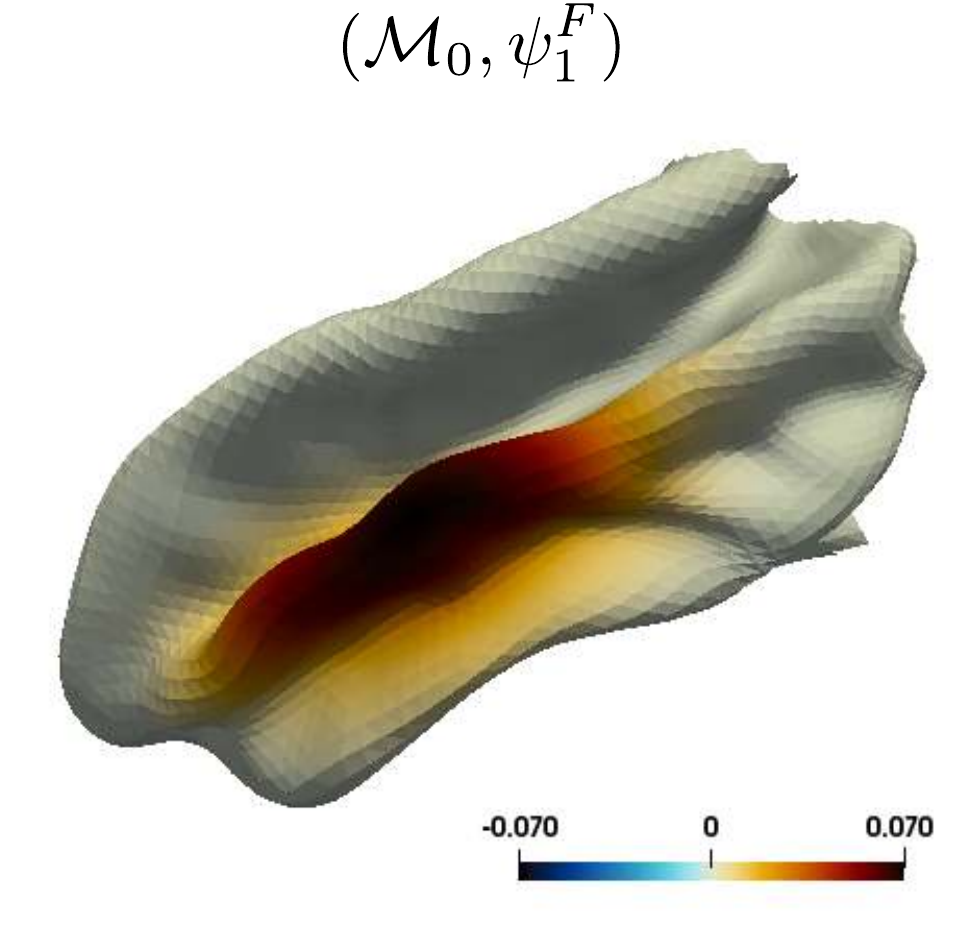}
\caption{On the left, a template of the temporal lobe $\Mtemp$ with an associated cortical thickness map $\mu$. On the right, the function $\psi_1^F$ used to generate subject-specific functional variability.}
\label{fig:temporal_lobe}
\end{figure}

\begin{figure}[ht]
\centering
\includegraphics[width=0.7\textwidth]{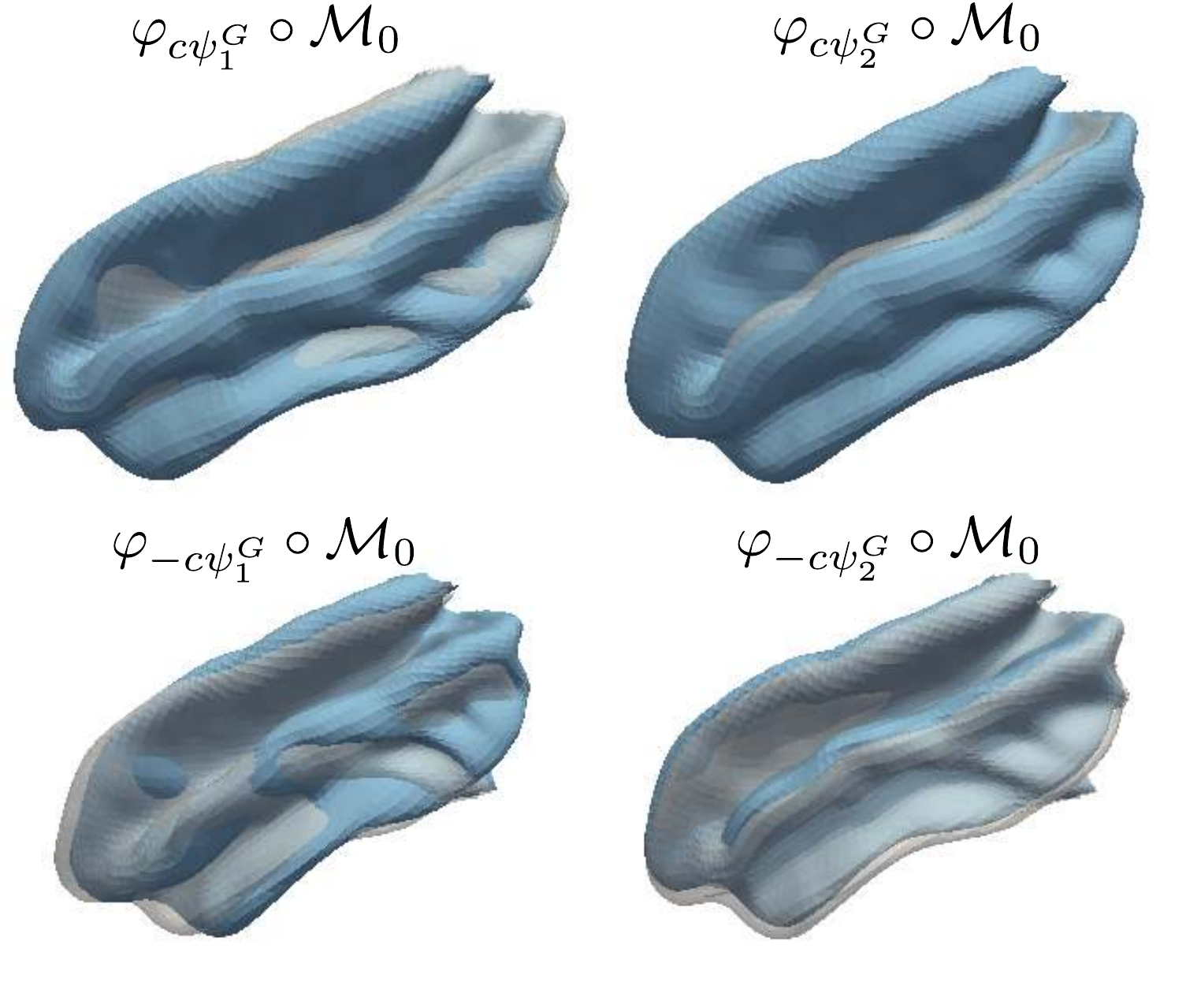}
\caption{From left to right, first and second geometric modes of variation of the generated \FoSs,  here visualized as $\varphi_{\pm c \psi_1^G} \circ \Mtemp, \varphi_{\pm c \psi_2^G} \circ \Mtemp$, where $c \in \R$ is a constant regulating the magnitude for visualization purposes.}
\label{fig:generation_geom}
\end{figure}

Thus, we denote with $\Mtemp$, the template temporal lobe shown in Figure~\ref{fig:temporal_lobe}.
We set the deformation operator $\varphi$ to be the diffeomorphic deformation operator introduced in Section~\ref*{sec:deformation_operator}. We then choose two orthonormal vector fields $\psi_1^G, \psi_2^G \in \calV$, visualized in Figure~\ref{fig:generation_geom} as the deformations $\varphi_{\pm c \psi_1^G}, \varphi_{\pm c \psi_2^G}$ applied to the template $\Mtemp$, where $c \in \R$ is a constant regulating the norm of the two orthonormal vector fields, for visualization purposes. The vector field $\psi_1^G$ encodes a change in the length of the temporal lobe, while the vector field $\psi_2^G$ encodes a change in the size of temporal lobe.

We set the mean function $\mu \in L^2(\Mtemp)$, to be the thickness maps in Figure~\ref{fig:generation_geom}, which is a sharpened version of the cross-sectional average thickness of 100 real subjects. Note that despite it being computed from real data, this plays the role of an unknown quantity of the model. Moreover, we introduce localized functional variability through the single mode of variation $\psi_1^F \in L^2(\Mtemp)$, this also visualized in Figure~\ref{fig:temporal_lobe}.

We then generate $n=50$ \FoSs\ $(\M_1, Y_1), \ldots, (\M_n, Y_n)$ by
\begin{align}\label{eq:data_generation}
\begin{cases}
\M_i &= \varphi_{a_{i1} \psi^G_1 + a_{i2} \psi^G_2} \circ \Mtemp,\\
X_i &= \mu + \delta a_{i2} \psi_1^F,\\
Y_i &= X_i \circ \varphi^{-1}_{a_{i1} \psi^G_1 + a_{i2} \psi^G_2},
\end{cases}
\end{align}
where $a_{i1}, a_{i2}$ are independent random variables distributed as $a_{il} \sim N (0, \sigma^2_l)$, with $\sigma_1 = 15$ and $\sigma_2 = 10$. The constant $\delta = 0.1$ determines the scale that relates variations in the functional term $\delta a_{i2} \psi_1^F$ and variations in the geometric term $a_{i2} \psi^G_2$. Finally, normally distributed noise with variance $\sigma=0.3$, is added to each node of the mesh where the function is observed. The generative model proposed here is a simplistic implementation of the one proposed in Model~\ref{eq:generation}, with $v_i = a_{i1} \psi^G_1 + a_{i2} \psi^G_2$ and $Z_i = a_{i2} \psi_1^F$.

The generative model (\ref{eq:data_generation}) seeks to reproduce a situation where the \FoSs\ have two modes of geometric variation. The first one is a mode of variation which is not correlated with a variation in the functions. The second one, which encodes the size of the temporal temporal lobe, has an effect on the function, formalized with a linear relation between the scores of the second geometric mode of variation $\psi_2^G$ and the scores of the functional mode of variation $\psi_1^F$. The generated \FoSs\ are such that larger temporal lobe have larger cortical thickness in proximity of the central gyrus of the cerebral cortex, independently of the first geometric mode of variation. We hope to recover this relation through the approximation pipeline introduced in Section~\ref*{sec:estimation_framework}.

\begin{figure}[!htb]
\centering
\includegraphics[width=0.28\textwidth]{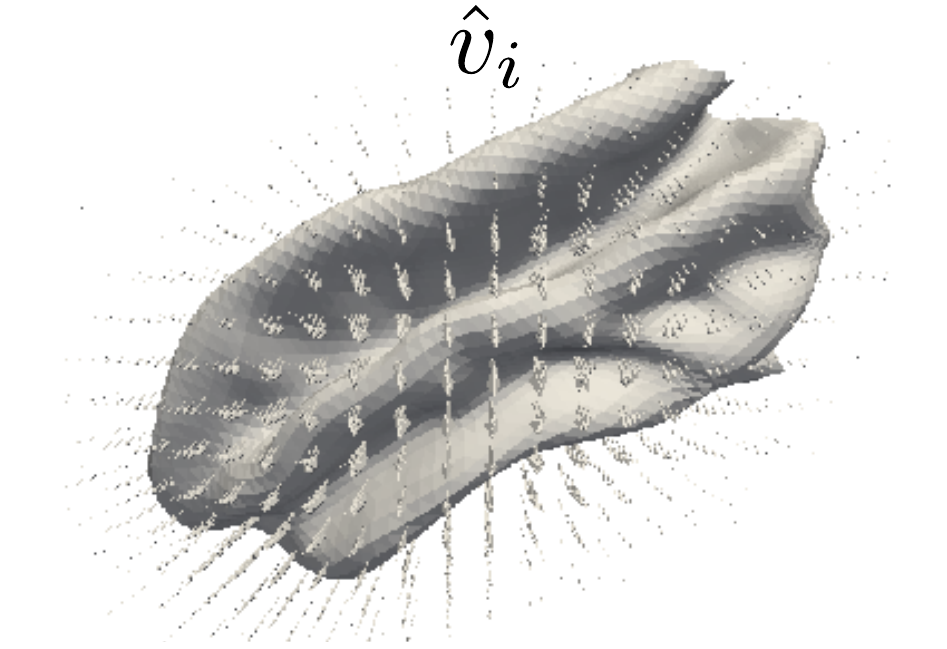}%
\includegraphics[width=0.72\textwidth]{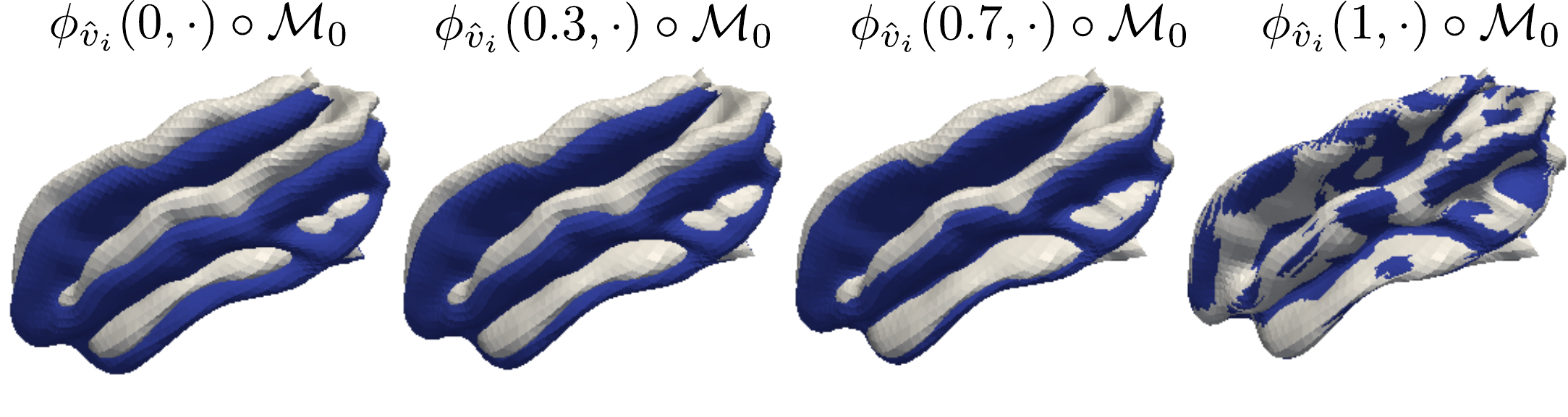}
\caption[]{On the left, the template $\Mtemp$ with an estimated vector field $\hat{v}_i \in \calV$ generating the diffeomorphic deformation $\varphi_{\hat{v}_i}$ that registers the template to the $i$th subject surface. Next, the evolution of the flow generating the diffeomorphic deformations $\phi_{\hat{v}_i}(t, \cdot)$ through the ODE (\ref{eq:ODE}), which registers the template to the target surface at time $t=1$.}
\label{fig:registration}
\end{figure}

\begin{figure}[!htb]
\centering
\includegraphics[width=1\textwidth]{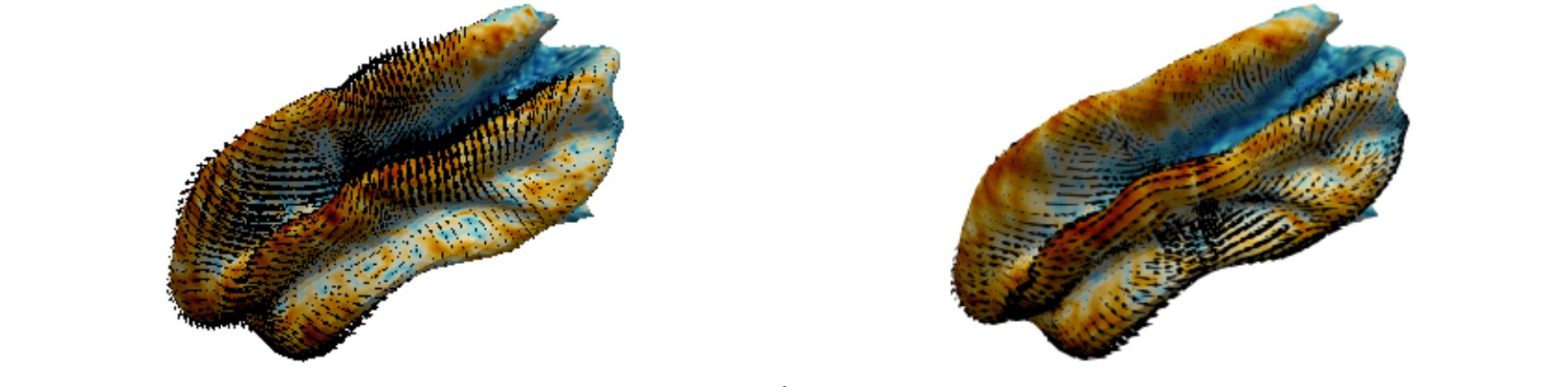}%
\caption[]{Two vector fields estimated from the functional registration algorithm, generating, for two different subjects, the flow which aligns two different functions to the cross-sectional mean function.}
\label{fig:registration_fun_vf}
\end{figure}

In particular, we perform non-landmarked diffeomorphic registration of the template to the single surfaces, resulting in the estimated vector fields $\{\hat{v}_i: i=1, \ldots,n \}$. The $i$th vector field $\hat{v}_i$ is such that $\varphi_{\hat{v}_i} \circ \Mtemp$ resembles the geometry $\M_i$ of the $i$th \FoS, with $\varphi$ the diffeomorphic deformation operator. In Figure~\ref{fig:registration}, we show an estimated vector field $\hat{v}_i \in \calV$ and the ODE's (\ref{eq:ODE}) flow $\phi_{\hat{v}_i}(t, \cdot)$, generated from the estimated vector field, which deforms the template to match the target.

The estimated diffeomorphic deformations $\{\varphi_{\hat{v}_i} = \phi_{\hat{v}_i}(1, \cdot)\}$ are then used to transport the functions $\{Y_i\}$ on the template surface, thus leading to the estimates $\{\hat{X}_i\}$. Subsequently, the cross-sectional mean map of $\{\hat{X}_i\}$ is computed and each function $\hat{X}_i$ is iteratively registered to it through the functional registration algorithm presented in Section~\ref*{sec:functional_registration_brief}. In Figure~\ref{fig:registration_fun_vf}, we show the template surface, with the tangential vector fields that generate the deformations that align two different functions to the cross-sectional mean function.

\begin{figure}[!htb]
\centering
\includegraphics[width=1\textwidth]{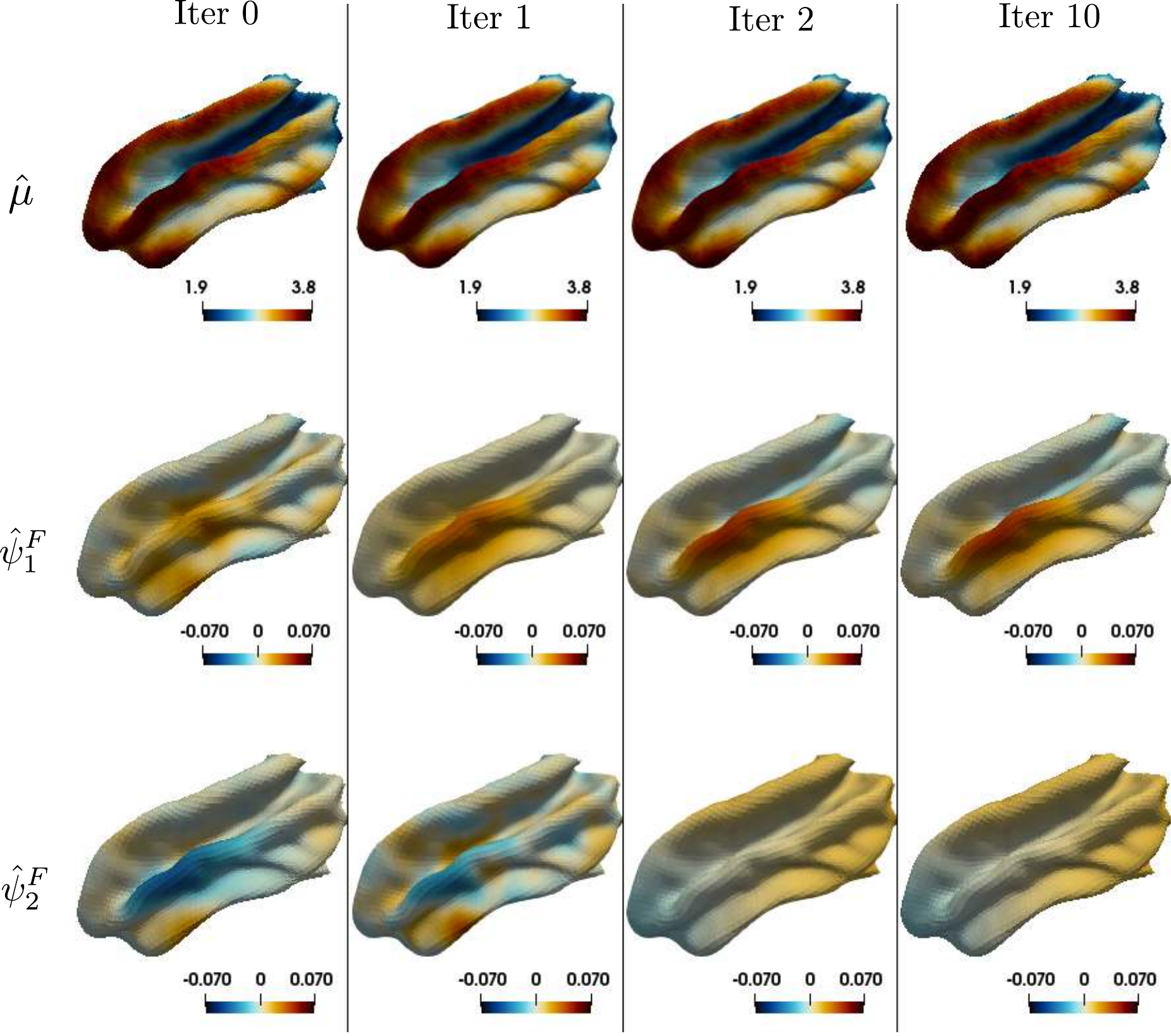}
\caption[]{From left to right, the mean and first two functional PC functions estimates of the functions, computed after $0,1,2$ and $10$ iterations of the functional registration algorithm.}
\label{fig:fPCA}
\end{figure}

At each iteration of the functional registration algorithm, the cross-sectional mean and the first 2 functional PCs, from the functionally aligned versions of $\{\hat{X}_i\}$, are computed. The results are shown in Figure~\ref{fig:fPCA}. We can notice that while the cross-sectional mean does not change from iteration to iteration of the functional registration algorithm, the estimates of the PC functions do. In particular, the first PC function is supposed to capture $\psi_1^F$. However, where no functional registration is applied, the first estimated PC component is a mix of the $\psi_1^F$ and fictitious variability due to misalignment, while the second PC function is a flat and corrupted version of $\psi_1^F$. After only one iteration of the functional registration algorithm, the estimated first PC function starts resembling the shape of $\psi_1^F$, shifting the misalignment component to the second PC function. With the subsequent iteration the first estimated PC function becomes a sharper estimation of $\psi_1^F$, while the misalignment component disappears also from the second component, in favour of a flat PC function, which is a regularized PC function of the noise added to the functions.

\begin{figure}[!htb]
\centering
\includegraphics[width=0.7\textwidth]{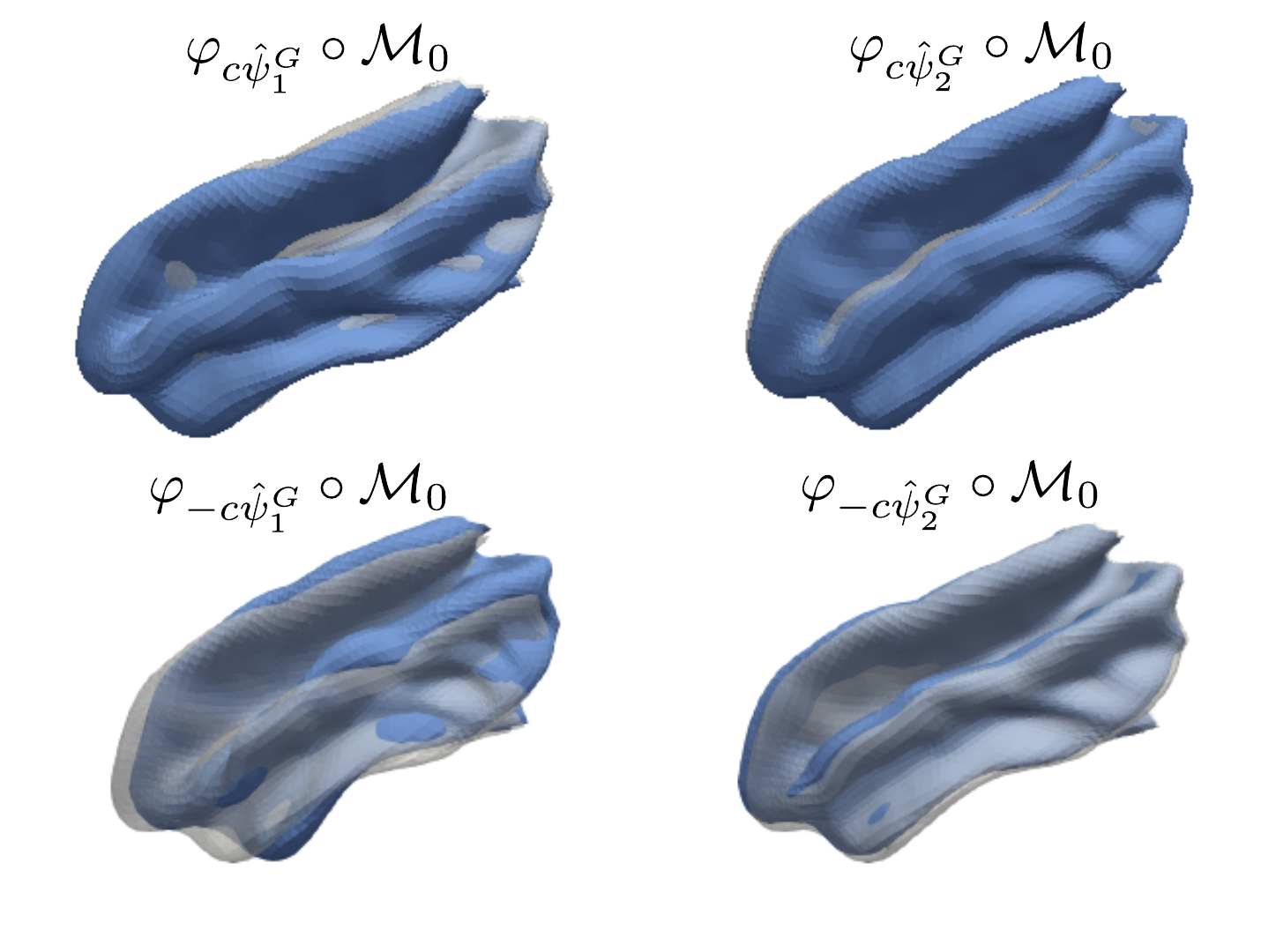}%
\caption[]{From left to right, the first two geometric PC functions computed on the space of initial vector fields. These are visualized as $\varphi_{\pm c \hat{\psi}^G_j} \circ \Mtemp$, where $\hat{\psi}^G_j$ is the estimated $j$th geometric PC function.}
\label{fig:tPCA}
\end{figure}

Subsequently, we perform fPCA on the estimated vector fields $\{\hat{v}_i\}$ representing the overall deformation, due to both geometric and functional registration. In Figure~\ref{fig:tPCA} we show the estimated main modes of variation before the functional registration has been applied. By comparison with Figure~\ref{fig:generation_geom}, we can see that the first two PCs capture the main geometric modes of variations introduced in the generative process of the \FoSs. The estimated geometric PC function do not change, in a visible manner, from iteration to iteration of the functional registration algorithm, because the functional registration brings only small deformations.

\begin{figure}[!htb]
\centering
\includegraphics[width=1\textwidth]{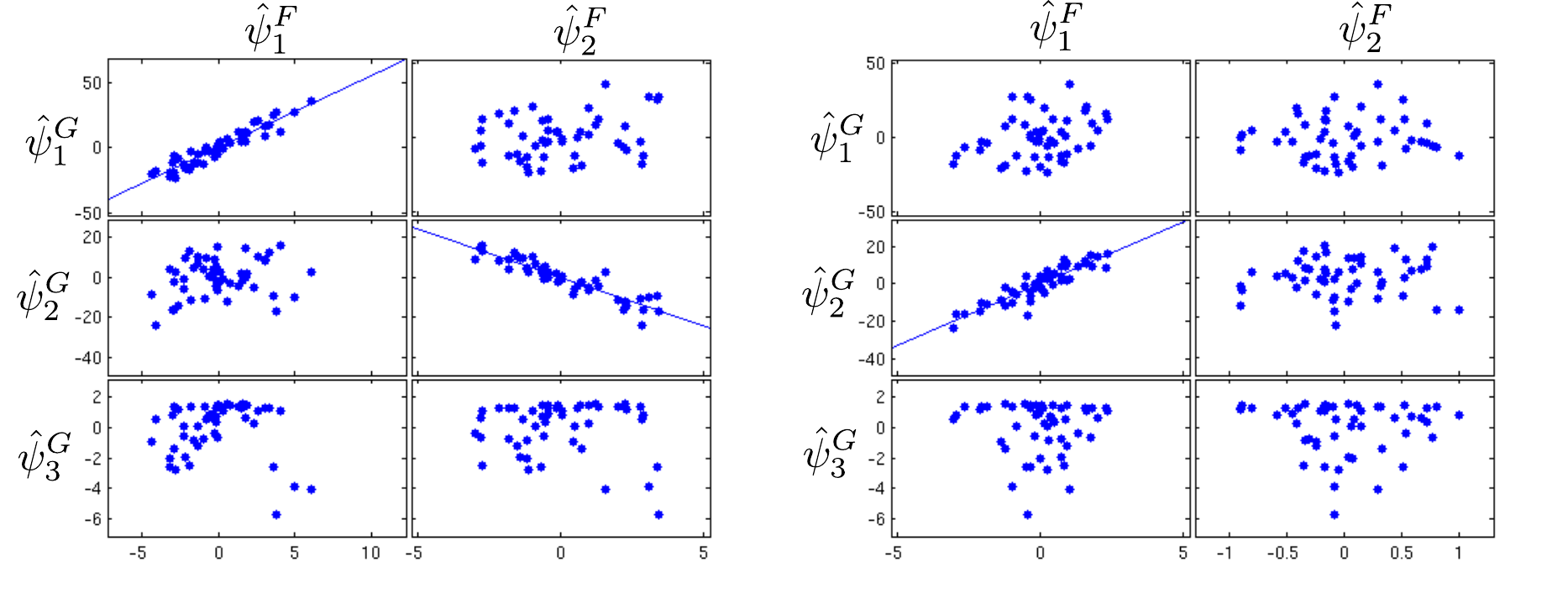}%
\caption[]{From left to right, scatter plots of the scores obtained from the fPCA on the function and the geometric fPCA, respectively without and with functional registration. After functional registration, these show only the linear dependence imposed between the first PC function on the functions $\psi_1^F$ and the second geometric PC function $\psi_2^G$. Without functional registration, also the spurious PC function, due to misalignment, is correlated with the first geometric PC function.}
\label{fig:regressionscores}
\end{figure}

We finally plot, in Figure~\ref{fig:regressionscores}, the scores associated to the PCs describing the geometric variability and those describing the functional variability, for the estimated quantity without functional registration and after seven iterations of the functional registration. Note that without performing functional registration, not only is the first functional mode of variation a spurious version of the true underlying component, but this is also correlated to the geometric mode of variations, which might lead to misleading conclusions. Functional registration removes from the first PC the misalignment effect, bringing to light the true underlying linear dependence between the functional mode of variation and the second geometric mode of variation.

In practice, the above procedure is particularly useful if the discovered PCs have biological interpretations. However, in practice, the discovered PCs tend to vary, depending for instance on the pre-registration method applied or on the scalar product adopted to impose orthogonality between the PC functions. For these reasons, if the aim is to study the relation between geometry and function, we advocate CCA (see Section~\ref*{sec:geo-fun_emp}).
\begin{figure}[!htb]
\centering
\includegraphics[width=1\textwidth]{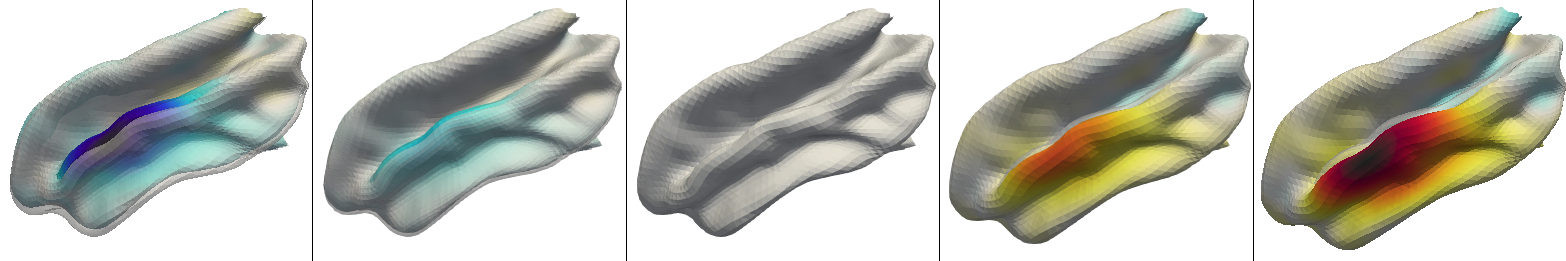}
\caption[]{First main mode of co-variation of geometric and functional components of the CCA analysis, representing the most correlated linear combinations of the first five geometric modes of variation and first three functional modes of variation. From left to right, this is visualized by plotting the \FoS\ in (\ref{eq:CCA_sim}) for a sequence of constants $c$.}
\label{fig:CCA}
\end{figure}
We perform a CCA on the estimated scores of the geometric and functional variability, after seven iterations of the functional registration algorithm. In detail, we construct a $n \times 3$ matrix $\mathbb{X}^F$ with the scores of the first three components of the fPCA applied to the functions. Moreover, we construct a $n \times 5$ matrix $\mathbb{X}^G$ with the scores of the first five components of the fPCA applied to the deformations. The $l$th canonical correlation component is the pair of vectors $\hat{\vect{w}}^{F,l} \in \R^3$ and $\hat{\vect{w}}^{G,l} \in \R^5$. The resulting main mode of co-variation $(\hat{\vect{w}}^{G}, \hat{\vect{w}}^{F}) = (\hat{\vect{w}}^{G,1}, \hat{\vect{w}}^{F,1})$ is visualized in Figure~\ref{fig:CCA} as
\begin{align*}
\begin{cases}
\begin{split}
\M_{\text{CCA}} &= \varphi_{c \; \hat{\psi}^G_{\text{CCA}}} \circ \Mtemp,\\
Y_{\text{CCA}} &= c \; \hat{\psi}^F_{\text{CCA}} \circ \varphi^{-1}{c \;  \hat{\psi}^G_{\text{CCA}}},
\end{split}
\end{cases}
\end{align*}
with $\hat{\psi}^G_{\text{CCA}} = \sum_{j=1}^3 \hat{w}^{G}_j \hat{\psi}^G_j$ and $\hat{\psi}^F_{\text{CCA}} = \sum_{j=1}^5 \hat{w}^{F}_j \hat{\psi}^F_j$, where $(\hat{\psi}^G_j)$ and $(\hat{\psi}^F_j)$ are the estimated functional and geometric PC functions. $c \in \R$ is a constant varied for visualization purposes in an interval containing $0$. As we can see in Figure~\ref{fig:CCA}, the dependence between the magnitude and the thickening of the function is captured.

Moreover, we test for the statistical significance of the obtained modes of co-variation. Specifically, we test the hypotheses
\begin{equation}\label{eq:test_CCA}
H_0^l: \hat{\rho}_1 \neq 0, \hat{\rho}_2 \neq 0, \ldots, \ldots, \hat{\rho}_l \neq 0, \hat{\rho}_{l+1} = \ldots = 0,
\end{equation}
with $\hat{\rho}_l = \text{corr}(\mathbb{X}^G \hat{\vect{w}}^{G,l}, \mathbb{X}^F \hat{\vect{w}}^{F,l})$. According to a likelihood ratio test, with the Bartlett $\chi^2$ approximation of the test statistic distribution \citep[see][Chapter~10.6]{Johnson2007}, only the sample correlation between the first canonical correlation variables, i.e. $\mathbb{X}^G \hat{\vect{w}}^{G,1}$ and $\mathbb{X}^F \hat{\vect{w}}^{F,1}$, is significantly different from zero (p-value $5\mathrm{e} -19$), while for $l=2,3$ we get p-values $0.7759$ and $0.9587$ respectively.

\section{Application}\label{sec:application}
The publicly available data set considered in this work has been collected by the Human Connectome Project Consortium \citep[HCP,][]{vanessen2012}, with the ultimate goal of elucidating the understanding of the brain functions, by collecting multi-modal neuroimaging data such as structural scans, resting-state and task-based functional MRI scans, and diffusion-weighted MRI scans from a large number of healthy volunteers.
A minimal preprocessing pipeline have been applied to the dataset \citep{glasser2013}.

\subsection{Preprocessing}
A 3D structural MRI scan has been performed for each individual, returning a 3D image describing the internal structure of the brain. A slice of the 3D image is shown on the left panel of Figure~\ref{fig:HCP_extraction}. The cerebral cortex is the outermost layer of the brain, mostly consisting of neuronal cell bodies. With automatic segmentation techniques, it is possible to separate the cerebral cortex from the other parts of the brain. Subsequently the two surfaces enclosing the cerebral cortex can be computed.
The inner surface represents the boundary between the cerebral cortex and the white matter (second panel in Figure~\ref{fig:HCP_extraction}), while the outer surface corresponds to the boundary between the cerebral cortex and the cerebrospinal fluid (fourth panel in Figure~\ref{fig:HCP_extraction}). The resolution of the MRI images (0.7 mm isotropic, in this study) and the effectiveness of the segmentation algorithm determine the level of details at which such surfaces can be reconstructed.

The geometry of the cerebral cortex is generally represented by the mid-thickness surface, which is the surface fitting the middle-points of the inner and outer surfaces, an example of which is shown on the third panel of Figure~\ref{fig:HCP_extraction}. Thus, it is natural to expect the resulting surfaces to have wider sulci and thinner gyri than what we could observe from a picture of the brain surface. Moreover, the mid-thickness surface can be equipped with a function representing the thickness of the cerebral cortex, computed from the inner and outer surface, as described in \cite{Fischl2000}. A comparison of the various methods for the cerebral cortex thickness estimation can be found in \cite{Lerch2005}. In Figure~\ref{fig:cortical_surfaces}, we show the reconstructed (mid-thickness) surfaces of the left hemisphere of 3 different subjects with the associated cerebral cortex thickness maps. Each surface is represented by a 32K nodes mesh, and at each node of the mesh an evaluation of the function is available.
\begin{figure}[!htb]
\centering
\includegraphics[width=0.7\textwidth]{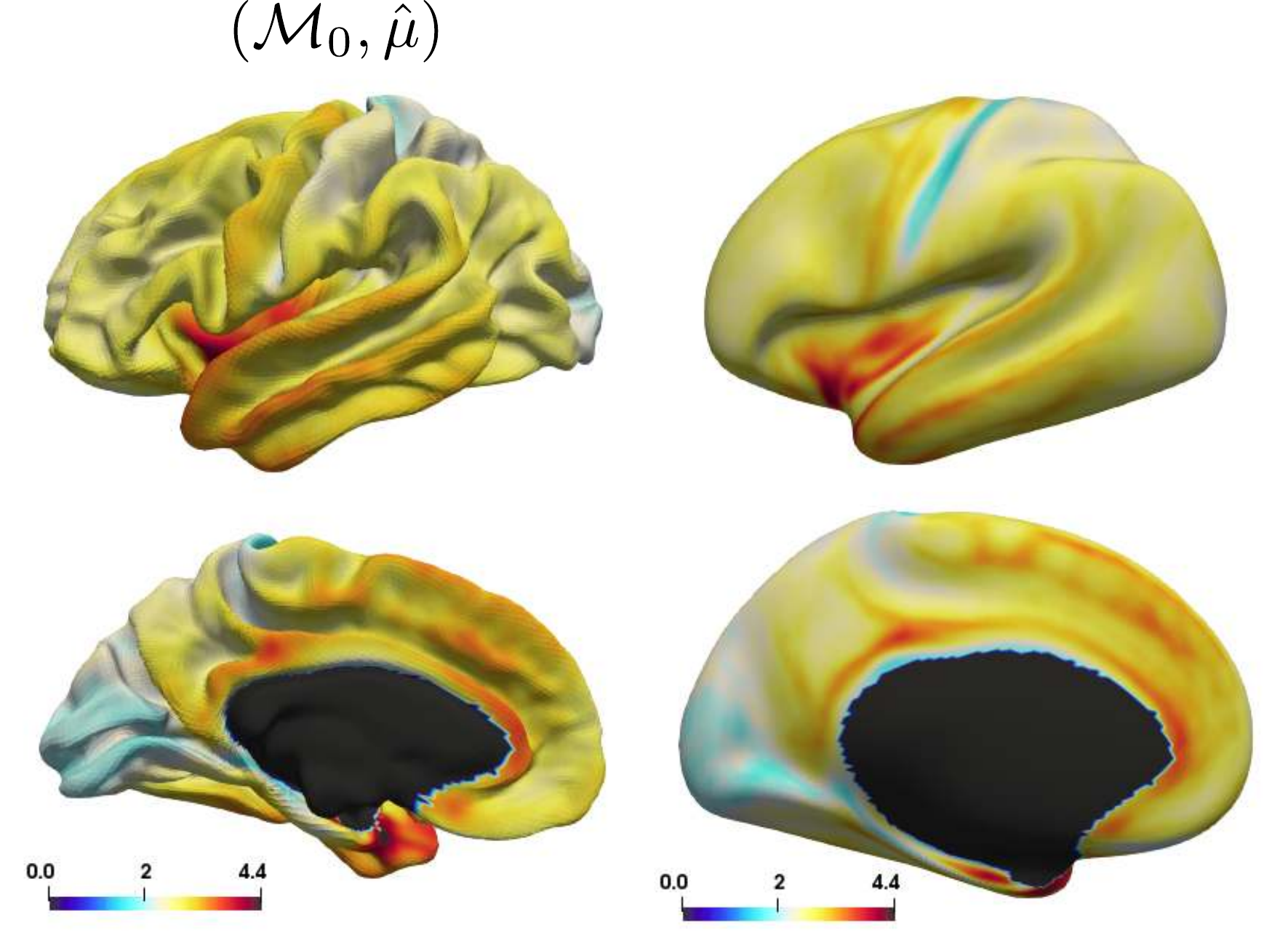}
\caption[]{On the left, the Conte69 template, used as a template surface for the registration of the individual surfaces. This is equipped with the cross-sectional mean function computed post-geometric registration. On the right, the cross-sectional mean function visualized on an inflated version of the template.}
\label{fig:HCP_mean}
\end{figure}

The mid-thickness surfaces of the collected cohort are pre-registered to the Conte69 template, on the left in Figure~\ref{fig:HCP_mean}, through a surface-based registration algorithm driven by geometric features that describe measures of cortical shape folding, such as sulcal depth or local curvature \citep{Fischl1999a,glasser2013}. Registrations are ensured to be one-to-one by introducing, in the objective function, a term related to the metric distortion of the registration maps and a term that enforces the positivity of the signed areas of the triangles on the surfaces \citep[see][for details]{Fischl1999a}. Such a procedure defines a one-to-one correspondence between the 32K nodes of the template and the 32K nodes of each of the mid-thickness surfaces, which can be regarded as a set of 32K landmarks.

\subsection{Analysis}
The relation between geometric features of the brain has raised great interest in the recent years, since it can potentially help us understand the principles underlying brain development. Classically, these studies have been confined to correlation studies on variables summarizing particular geometric features. For instance, in \cite{Im2008}, for each subject, the average cortical volume and absolute mean curvature, among other, are computed. This set of real variables are then compared to the average cerebral cortex thickness computed on each subject. Moreover, a more localized analysis is performed by parcellating each cortical surface in the 4 lobes.  Subsequently, the analysis is performed independently on each of lobe. However, there are two limitations of such approach. Firstly, the description of the geometric properties through summary statistics is in general incomplete. Secondly, the parcellation of the cortical surfaces determines a priori which areas of the cortical surface can have a different behaviour.

\begin{figure}[!htb]
\centering
\includegraphics[width=1\textwidth]{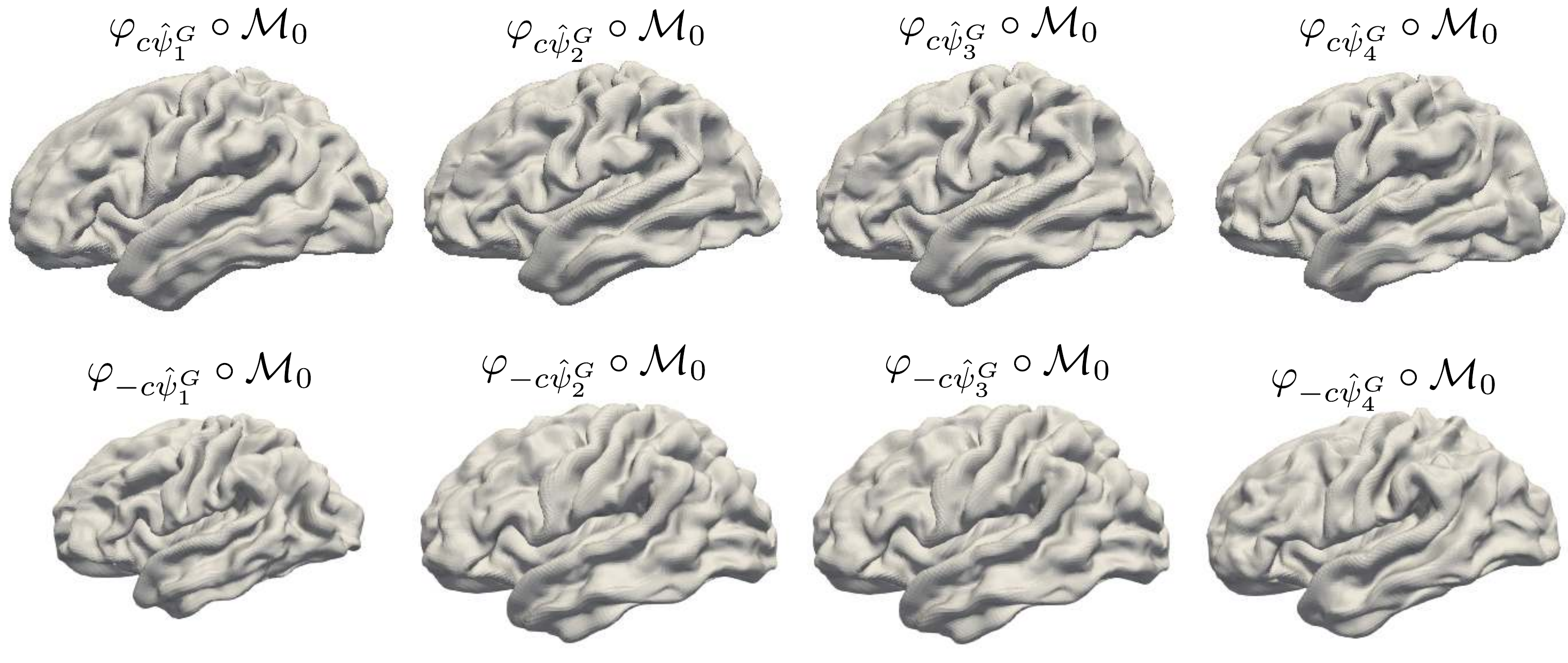}
\caption[]{From left to right, the first four geometric PC functions,  computed on the space of initial vector fields. These are visualized as $\varphi_{\pm c \hat{\psi}^G_j} \circ \Mtemp$, where $\hat{\psi}^G_j$ is the $j$th geometric PC function.}
\label{fig:HCP_tPCA}
\end{figure}

The fact that a geometric registration has already been performed on the HCP data, without relying on the diffeomorphic registration framework in Section~\ref{sec:geometric_registration_brief}, is not in contrast with the proposed analysis. In fact, diffeomorphic-like constraints can be imposed in many different ways when it comes to the estimation of registration maps. However, if the aim is the estimation of a low-dimensional subspace of the diffeomorphic space, these alternative approaches cannot be extended to this more general problem. For this reason, we use the landmarks defined by the pre-processing geometric registration to estimate the vector fields that represent such registrations and then perform fPCA on the estimated vector fields, as described in Section~\ref{sec:fPCA_brief}. The estimated first four geometric PCs are shown in Figure~\ref{fig:HCP_tPCA}. Not surprisingly, they are mostly related to the size of the brain or the size of sub-parts of the brain.

\begin{figure}[!htb]
\centering
\includegraphics[width=1\textwidth]{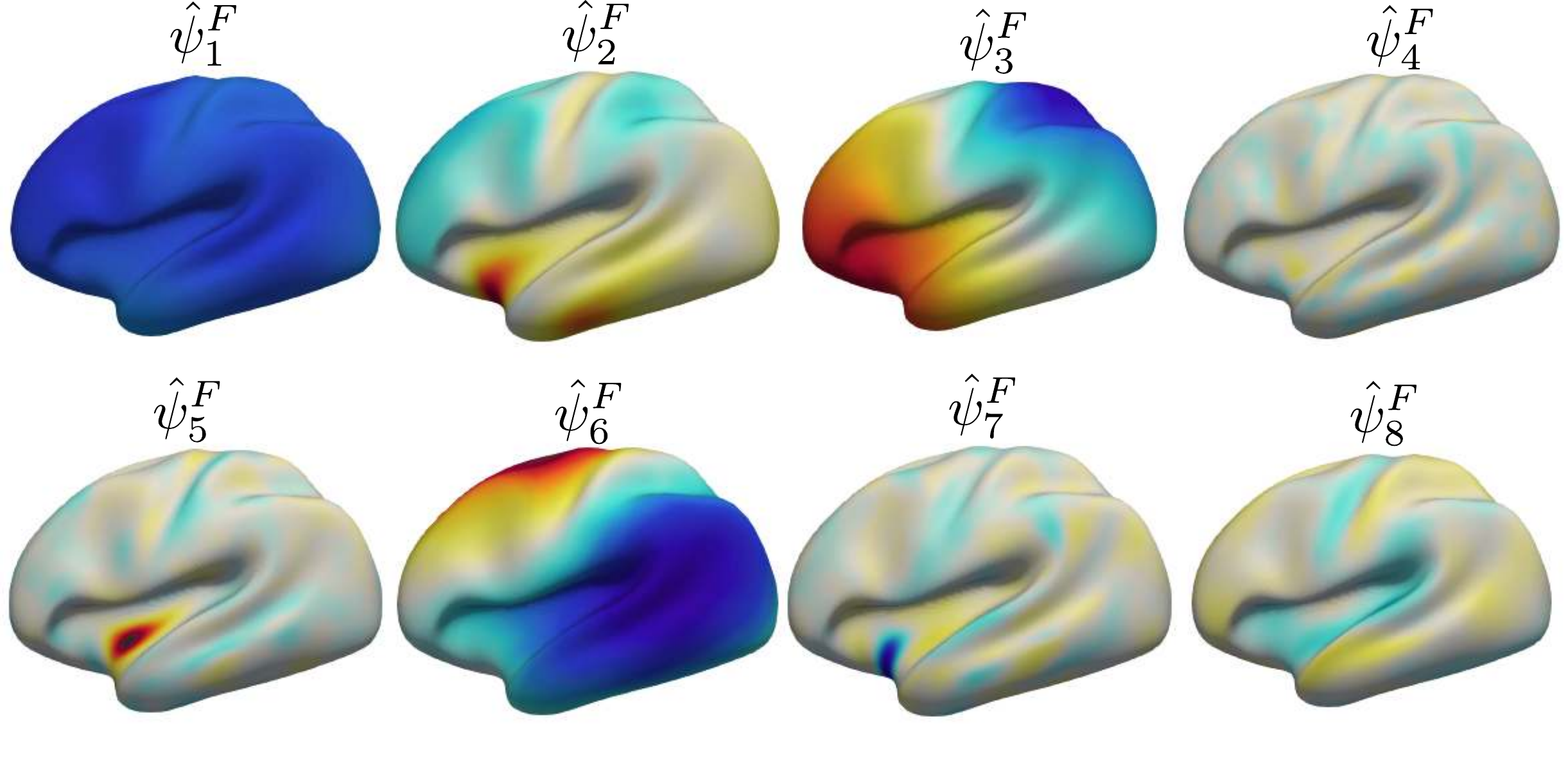}\\
\includegraphics[width=1\textwidth]{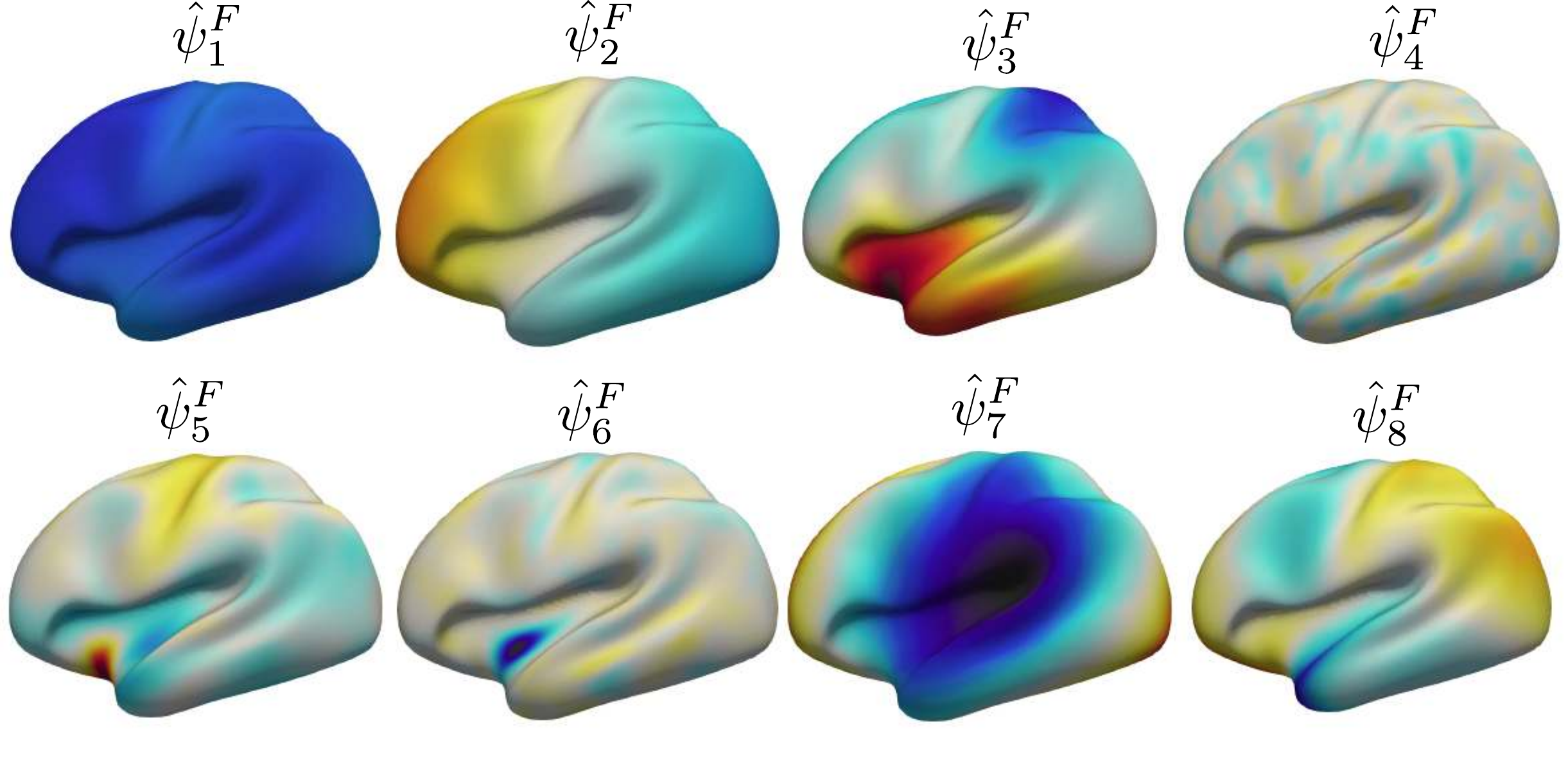}
\caption[]{Results of the fPCA on the functions. On the top two rows, the first eight functional modes of variation computed without performing functional registration. On the bottom two rows, the first eight functional modes of variation computed after performing functional registration. These are shown on an inflated version of the template to easy their visualization.}
\label{fig:HCP_fPCA}
\end{figure}

We then perform fPCA on the functions registered on the Conte69 template. The results are shown in the top two rows of Figure~\ref{fig:HCP_fPCA}. Subsequently, we perform functional registration of the functions on the Conte69 template and recompute the functional modes of variation at each iteration. In the bottom two rows of Figure~\ref{fig:HCP_fPCA} we show the PC functions after 2 iterations of the functional registration algorithm.

\begin{figure}[!htb]
\centering
\includegraphics[width=1\textwidth]{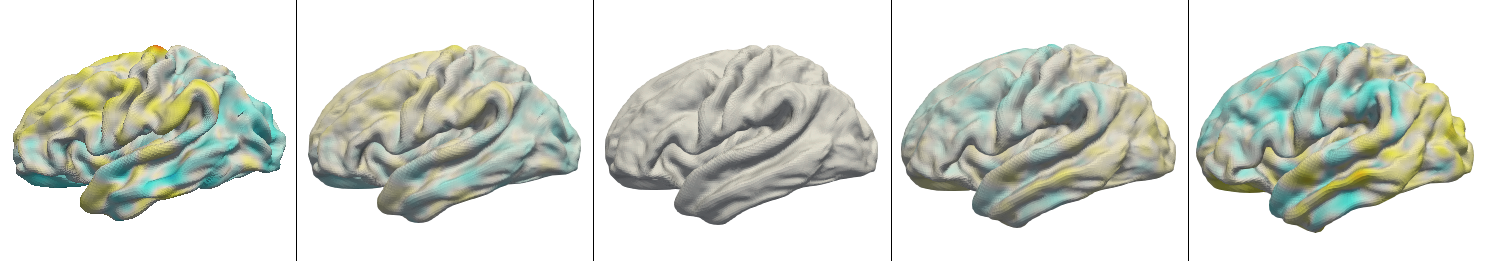}
\includegraphics[width=1\textwidth]{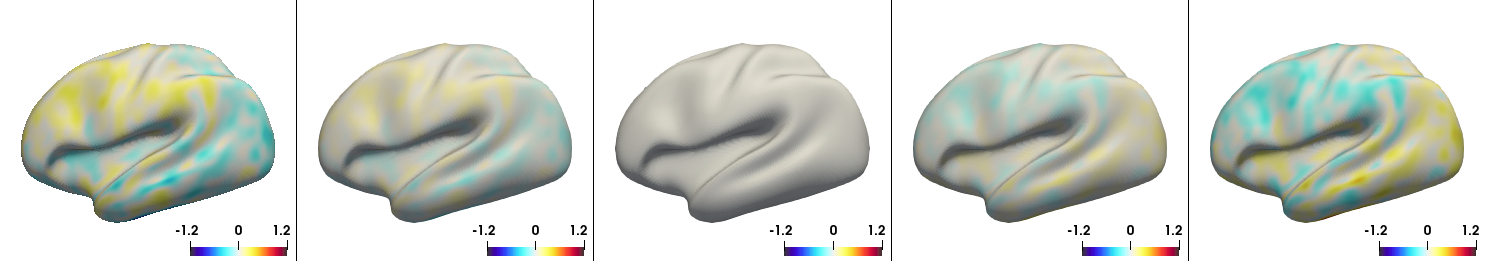}
\caption[]{A representation of the first main mode of co-variation of the geometric and functional components of the CCA analysis, representing the most correlated linear combinations of the first eight geometric and functional PC functions. From left to right, this is visualized by plotting the \FoS\ in (\ref{eq:CCA_sim}) for a sequence of constants $c$.}
\label{fig:HCP_CCA_1}
\end{figure}

\begin{figure}[!htb]
\centering
\includegraphics[width=1\textwidth]{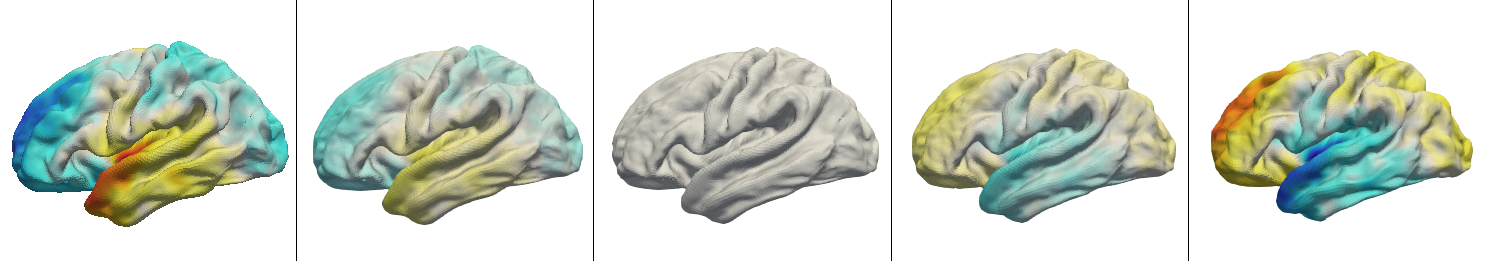}
\includegraphics[width=1\textwidth]{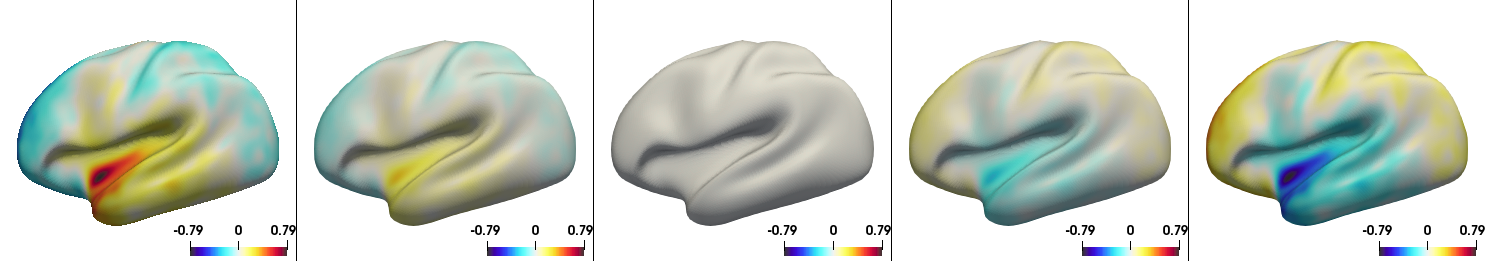}
\caption[]{A representation of the second main mode of co-variation of the geometric and functional components of the CCA analysis, representing the second most correlated linear combinations of the first eight geometric and functional PC functions. From left to right, this is visualized by plotting the \FoS\ in (\ref{eq:CCA_sim}) for a sequence of constants $c$.}
\label{fig:HCP_CCA_2}
\end{figure}

We finally perform a CCA on the first eight geometric and functional PC functions scores. The resulting first two main modes of co-variation, the only significant ones from the likelihood ratio test (\ref{eq:test_CCA}), are shown in Figures~\ref{fig:HCP_CCA_1}-\ref{fig:HCP_CCA_2}. From the left to the right panel of Figure~\ref{fig:HCP_CCA_1}, we can see the presence of a correlation between a decrease in thickness in the frontal lobe and an increase in size of the entire brain, while in the temporal lobe, an increase in thickness seems associated to an increase in size of the entire brain. Moreover, in the second main mode of co-variation a more localized phenomenon is captured in proximity of the high average cortical thickness area on the lateral sulcus (see Figure~\ref{fig:HCP_mean}), where an association between an increase in the cortical thickness and an increase in the size of the brain is suggested. Note that such local effect would have not been captured by a study confined to study individually each lobe of the brain, and such relation would have probably been ascribed to the entire lobe containing that area.

\section{Conclusions and Prospectives}\label{sec:conclusions}
In this paper, motivated by the analysis of neuroimaging data, we introduce a framework for the analysis of \FoSs. In particular, a statistical model describing the phenomenon is formulated, and the estimators of the unknown quantities of the model are introduced. The construction of such estimators is complicated by the necessity of the resulting estimates to lie in the non-linear subspace of `sensible' solutions, here taken to be deformations. Moreover, in such high dimensional setting, it is fundamental for the estimator to incorporate prior information on the geometry and the smoothness of the data, achieved by regularizing the estimates through differential operators. Motivated by simulation studies, we address the necessity of using the functional information to achieve a better registration, a well known fact in FDA, by introducing a novel diffeomorphic registration algorithm for functional data on a two-dimensional manifold.

While the main motivation of this paper was taken from a neuroimaging application into assessing the inherent variabilities of cortical thickness, the methodology has wider applications in medical imaging as a whole, where \FoS\ appear in cardiovascular (e.g. \citet{Huang2016}), muskuloskeletal (e.g. \citet{Treece2015}) and many other imaging areas. More generally, this methodology is an example of the use of differential operators as regularisers in statistics, a field where not only statistical but also numerical techniques are needed to facilitate solutions.

A future interesting aspect is the exploration of the applicability of the Optimal Transport framework to the registration problem, as suggested in \cite{Panaretos2016} in a discrete context, and its links with the diffeomorphic deformation framework. This is of potential interest in the surface registration framework, where we usually lack physical models that can describe the phenomena, and thus a `least action' approach could well be effective.

\section*{Acknowledgements}
The authors greatly appreciate the really useful comments of the AE and two referees, which helped considerably strengthen the paper. JA was supported by the Engineering and Physical Sciences Research Council (EP/K021672/2 and EP/N014588/1). EL was supported by the EPSRC grant EP/L016516/1.

\appendix
\appendixpage

\section{Geometric Diffeomorphic Registration}\label{app:diffeomorphic_registration}
Here we present the algorithmic details of the estimation framework introduced in Section~\ref*{sec:estimation_framework}.

The space of smooth vector fields $\calV$, in the geometric registration model (\ref{eq:minimization_geo}), is usually constructed as a RKHS \citep{Miller2015}. In detail, let $K_\calV:\R^3 \times \R^3 \ra \R^{3 \times 3}$ be a bounded symmetric positive definite function. $K_\calV$ is usually referred to as the kernel of $\calV$ and a typical choice for it is the Gaussian isotropic kernel, i.e. $K_\calV(x,y)= \exp(-\| x-y \|_2^2/ (2\sigma_{\calV}^2)) \Id_{3 \times 3}$, with $\Id_{3 \times 3}$ denoting a $3 \times 3$ identity matrix and $\sigma_{\calV}$ reflecting the rigidity of the space. Define the pre-Hilbert space $\calV_0 = \text{span} \{K_\calV(\cdot,x)\omega | x \in \R^3, \omega \in \R^3 \}$. Given $f,g \in \calV_0$ we can write them as $f = \sum_{i=1}^{N} K_\calV(\cdot,x_i) \omega_i$ and $g = \sum_{i=1}^{N} K_\calV(\cdot,y_i) z_i$. We thus define the inner product between $f$ and $g$ to be $\langle f,g \rangle_{\calV} = \sum_{i,j=1}^N \omega_i^T K_\calV(x_i,y_j) z_j$. The space $(\calV, \langle \cdot,\cdot \rangle_\calV)$, defined as the closure of $\calV_0$, is a (Reproducing Kernel) Hilbert space of smooth vector fields.

For modeling purposes, the time-variant vector-field $v_t$, introduced in Section~\ref*{sec:deformation_operator}, is assumed to be of the form \citep[see e.g.][]{Vaillant2004}
\begin{equation}
v_t (\cdot) = \sum_{k=1}^{k_g}K_\calV(\phi_v(t, c_k), \cdot) \alpha_k(t),
\end{equation}
for a set of control points $\{c_k: k=1,\ldots,k_g\} \subset \R^3$ and the auxiliary variables $\{\alpha_{k}(t): \R \ra \R^3\}$ called momenta of the deformation. The control points $\{c_k\}$ are commonly chosen to be the nodes of the triangulated representation of the surface to be deformed. $\phi_v$ denotes the solution of the ODE (\ref{eq:ODE}) given the time-variant vector field $\{v_t : t \in [0,1]\}$. The associated deformation energy is defined to be
\begin{equation}\label{eq:energy_vf}
\int_0^1 \|v_t\|^2_\calV = \int_0^1 \sum_{k,l=1}^{k_g} \alpha_k(t)^T K_\calV(\phi_v(t,c_k), \phi_v(t,c_l)) \alpha_l(t).
\end{equation}

Denoting with $\grad_1$ the gradient with respect to the first variable, the vector field $v_t$ generating geodesics, with respect to the energy term $\int_0^1 \|v_t\|^2_\calV$, can be characterized as the solution of the coupled ODE system, known as the EPDiff equation \citep{Miller2015}
\begin{align}\label{eq:hamiltonian}
\begin{cases}
\frac{\partial c_k(t)}{dt} &= \sum_{l=1}^{k_g} K_\calV(c_k(t),c_l(t))\alpha_l(t)\\
\frac{\partial \alpha_k(t)}{dt} &= -\frac{1}{2} \left( \sum_{l=1}^{k_g}\grad_1 K_\calV(c_k(t),c_l(t))\alpha_l(t)) \right)^T \alpha_k(t),
\end{cases}
\end{align}
for a set of initial conditions $\{\alpha_{k} = \alpha_k(0)\} \subset \R^3$, parameterizing the initial vector field $v_0$. This means that the energy minimizing vector fields, generating diffeomorphisms, can be determined by (\ref{eq:hamiltonian}) and fully controlled by the initial vector field
\[
v_0(\cdot)= \sum_{k=1}^{k_g} K_\calV(\cdot,c_k)\alpha_{k},
\]
parametrized in terms of the initial momentum vector $\{ \alpha_{k}: k=1,\ldots, k_g \}$. Moreover, along a geodesic path the instantaneous deformation energies $\|v_t\|_\calV$ are constant, meaning that the total deformation energy $\int_0^1 \|v_t\|^2_\calV dt$ can be equivalently represented by the initial deformation energy $\|v_0\|^2_\calV = \sum_{k,l} \alpha_{k} K_\calV(c_{k},c_{l})\alpha_{l}$.

Thanks to the finite dimensional representation underlying the element of the RKHS $\calV$, the minimization of (\ref{eq:minimization_geo}) can be cast in a finite dimensional setting and can be approached, for instance, with a gradient descent algorithm on the initial momentum vector parametrizing the initial velocity field \citep[see, among others,][]{Vaillant2004}.

The MATLAB toolkit fshapesTk (\href{https://github.com/fshapes/fshapesTk}{https://github.com/fshapes/fshapesTk}) offers an implementation of the described geometric registration algorithm, and its extension to the fshapes framework \citep{Charlier2017}.

\section{Registration of Functional Data on a two-dimensional manifold}\label{app:functional_reg}
Here we cover further details of the functional registration algorithm, for functional data whose domain is a two-dimensional manifold, introduced in Section~\ref*{sec:functional_registration_brief}. The main idea of the proposed algorithm is to perform functional registration as compositions of small diffeomorphisms, each parameterized by a stationary velocity field. This class of algorithms are also known as Diffeomorphic Demons algorithms \citep{Vercauteren2009}. Diffeomorphic Demons were originally introduced for functions on Euclidean domains and an extension to spherical domains has been proposed in \cite{Yeo2010}.  However, this extension exploits spherical vector spline interpolation theory and cannot be extended to a generic manifold. In the geometric registration problem, as detailed in Section~\ref{app:diffeomorphic_registration}, smoothness is imposed by controlling the norm $\| \cdot\|_\calV$ of the functional space. In fact, in $\R^3$, it is easy to define symmetric definite positive kernels from which we can straightforwardly define $\calV$ thanks to the RKHS machinery. This approach does not easily extend to non-linear domains such as $\Mtemp$.

For this reason, here we rely on a construction of the space of smooth vector fields $\calW$ based, instead, on the definition of a differential operator encoding smoothness, as done for instance in the planar 2D case in \cite{Beg2005}. However, in the planar 2D case a matrix operator for a vector field $u:\R^2 \ra \R^2$ can be defined as the isotropic Laplacian operator
\[
    \begin{bmatrix}
    \Delta  & 0 	 \\
    0 		& \Delta \\
\end{bmatrix},
\]
where $\Delta$ is the Laplacian operator for real valued functions.
The isotropic Laplacian applies the Laplacian operator component-wise to a vector field in $\R^2$, exploiting the fact that, in the Euclidean space $\R^2$ there is a global reference system. The introduction of an analogous operator for vector fields on a manifold is not straightforward for the main reason that nearby vectors live on different tangent spaces. The definition of such a coordinate independent operator for vector fields requires additional notions of Riemannian geometry.  In particular, we rely on the Bochner-Laplacian, which is used to enforce smoothness on the vector fields generating diffeomorphism on a manifold.

\subsection{Differential operators on tangent vectors}\label{app:bochner-laplacian}
Recall that we denote with $T_p \Mtemp$ the tangent space on the point $p \in \Mtemp$ and with $g_p$ the metric on $\Mtemp$. Moreover we denote with $T \Mtemp = \dot\bigcup_{p \in \Mtemp} T_p \Mtemp$ the tangent bundle, i.e. the disjoint union of tangent spaces. The space of smooth sections of the tangent bundle $T \Mtemp$, i.e. the space of smooth vector fields on $\Mtemp$, is denoted with $\Gamma (T\Mtemp)$.

Given a tangent vector $w \in T_p \Mtemp$, a vector field $u \in \Gamma (T\Mtemp)$ and a smooth function $f: \Mtemp \rightarrow \R$, an affine connection $\nabla$ is an operator such that $\nabla_w u \in T_p \Mtemp$. Moreover, it is linear in both $w$ and $u$ and is such that it satisfies the Leibniz rule, namely
\begin{equation*}
\nabla_w (fu) = df(w) u + f \nabla_w u.
\end{equation*}
For a manifold $\Mtemp$ embedded in a Euclidean space, by requiring that the affine connection $\nabla$ must preserve the metric and must be torsion free, we have that this can be uniquely determined. Under these hypotheses, $\nabla$ is called the Levi-Civita connection.
\begin{figure}[ht]
\centering
\includegraphics[width=.3\textwidth]{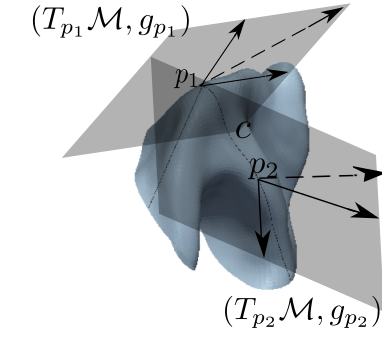}
\caption{The figure is a pictorial representation of the parallel transport of the striped arrow from $p_1$ to $p_2$. Note that because of the different reference systems in $p_1$ and $p_2$, expressing the vector as a linear combination of the basis element in $T_{p_2} \Mtemp$ with the same coefficients as in $T_{p_1} \Mtemp$ would yield to a different result.}
\label{fig:parallel_transport}
\end{figure}
In practice a connection defines a way to generalize parallel transport on a manifold. In fact, the parallel transport of a vector $u \in T_p \Mtemp$ along a curve $c$ can be defined as the collection of vectors along the curve $c$ such that $\nabla_{c'(s)} u = 0$, where $c'(s) \in T_{c(s)} \Mtemp$. A pictorial representation of this is given in Figure~\ref{fig:parallel_transport}. Finally, we can define the Bochner-Laplacian operator, of a smooth section of $v \in \Gamma(T \Mtemp)$,  as
\begin{equation}
\Delta_{BL} = \nabla^* \nabla
\end{equation}
where $\nabla^*$ is the $L^2$ adjoint of $\nabla$.

\subsubsection{Functional Registration Model}\label{app:functional_reg_model}
Let now $M,F: \Mtemp \ra \R$ be respectively a `moving' and `fixed' image. We recall here the objective function of the functional registration model (\ref{eq:diff-nonlinear}), in terms of $M$ and $F$:
\begin{equation}\label{eq:diff-nonlinear_app}
E_{\Mtemp}(u) = \sum_{j=1}^S \big (F(p_j) - M \circ s \circ \phi_u (p_j) \big)^2 +  \lambda\|\Delta_{BL} u\|^2_{L^2(T\Mtemp)},
\end{equation}
with $\{ p_j, j=1,\ldots,S\} \subset \Mtemp$ the set of control points on the template and $\phi_u$ denoting the solution of the ODE (\ref{eq:ODE-stationary}) for the vector field $u$, at time $t=1$.

\begin{figure}[!htb]
\centering
\includegraphics[width=0.4\textwidth]{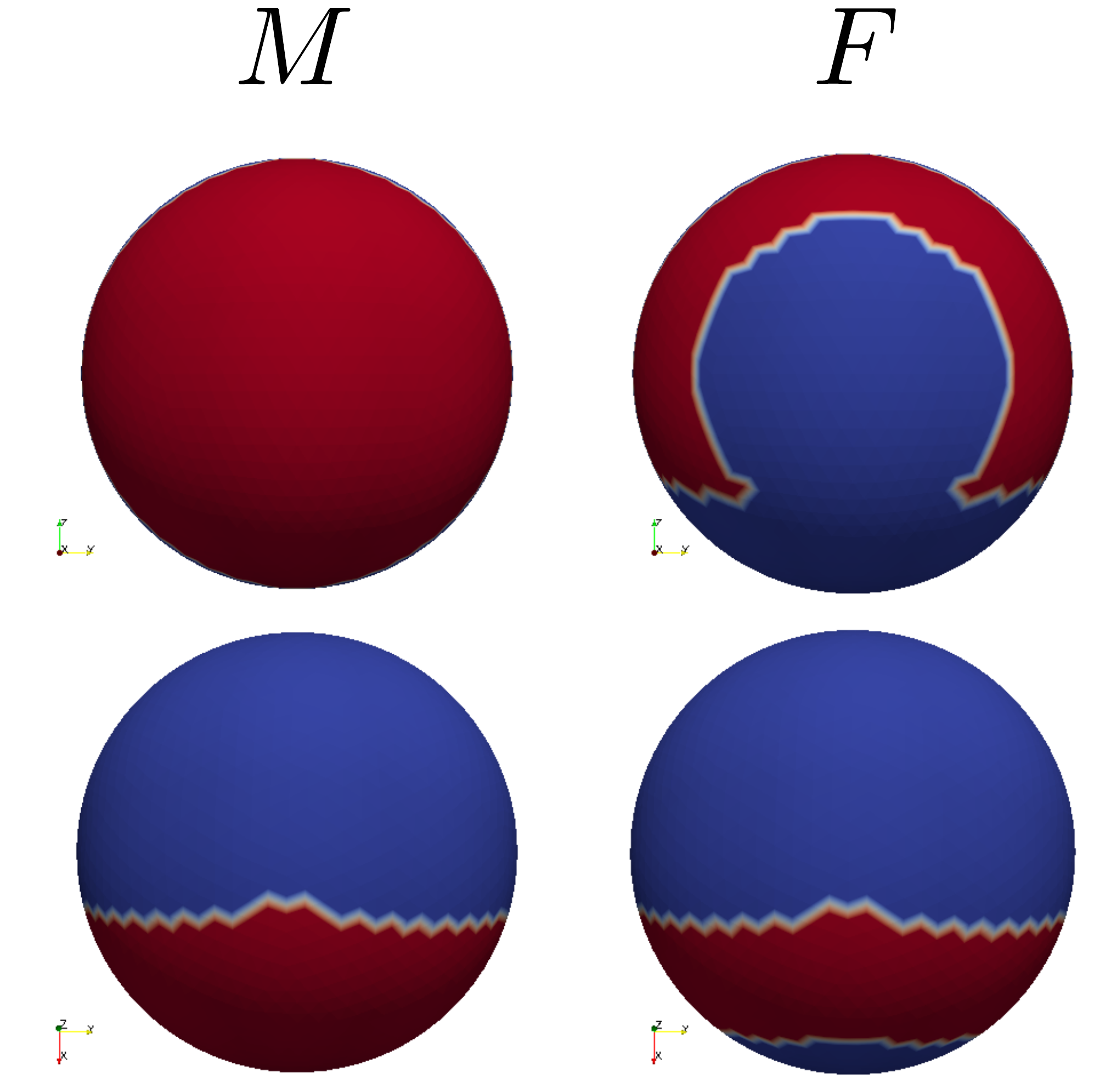}
\caption{On the left two views of a semi-circle image on the unit sphere, representing the moving image $M$, while on the right two views of a C-shaped image on the unitary sphere, representing the fixed image $F$.}
\label{fig:fun_reg_images}
\end{figure}

The term $M \circ s \circ \phi_{u}$ is then linearized with respect to $u$. This results in the approximation
\[
M \circ s \circ \phi_{u} \approx M \circ s + L_{u},
\]
where $L_{u}$ is a first order approximation of $M \circ s \circ \phi_{u} - M \circ s$. In practice $L_{u}$ is chosen to be of the form
\[
L_{u}(p) = g_{p}(J(p),u(p)), \qquad p \in \Mtemp,
\]
with $J(p) \in T_p\Mtemp$ for all $p \in \Mtemp$. Two classical choices for $J$, in the planar case, are $J = \nabla_D(M \circ s)$ and $J = \frac{1}{2} (\nabla_D(M \circ s) + \nabla_D(F))$ \citep{Vercauteren2009}, where $\nabla_D$ denotes a discrete estimate of the gradient. Plugging the linearized term in (\ref{eq:diff-nonlinear_app}) we obtain the objective function
\begin{equation}\label{eq:diff-linear_app}
E_{\Mtemp}(u) = \sum_{j=1}^S \big (F(p_j) - (M \circ s)(p_j) - g_{p_j}(J(p_j),u(p_j)) \big)^2 +  \lambda\|\Delta_{BL} u\|^2_{L^2(T\Mtemp)}.
\end{equation}

The minimization of (\ref{eq:diff-nonlinear_app}) can be achieved by iteratively minimizing the associated problem (\ref{eq:diff-linear_app}) and updating the current deformation $s$ with $s \ra s \circ \phi_u$, with $\phi_u$ denoting the solution of the ODE (\ref{eq:ODE-stationary}) at time $t=1$.

\subsection{Problem reformulation}\label{app:VF-FE}
To minimize the objective function in (\ref{eq:diff-linear_app}) we opt for a finite elements discretization approach. Finite element discretization has been previously applied to the discretization of FDA problems on manifolds, for instance, in \cite{SSRM1} and \cite{Lila2016}. Here, we extend the methodology to the estimation of smooth vector fields on a generic two-dimensional manifold. To this end, we first reformulate the minimization of (\ref{eq:diff-linear_app}) in terms of the Euler-Lagrange equation associated to this minima problem.

Define now the space of smooth vector fields on the template to be $\calW = \{ u \in L^2(T\Mtemp) | \Delta_{BL} u \in L^2(T\Mtemp) \}$.  Let the vector field $u \in \calW$, in the functional (\ref{eq:diff-linear_app}), be perturbed by an $\epsilon$ amount along the arbitrary direction $\varphi \in \calW$. The minimization problem is reformulated by imposing the Gateaux derivative $\partial_\varphi E_\Mtemp(u)$ of the energy functional to be $0$ for all $\varphi \in \calW$. 

This leads to the problem reformulation: find $\hat{u} \in \calW$ such that
\begin{align}\label{eq:euler-lagrange-VF}
\begin{split}
\sum_{j=1}^S g_{p_j}(\varphi(p_j), J(p_j)) g_{p_j}(\hat{u}(p_j),J(p_j)) + \lambda \langle \Delta_{BL} \varphi, \Delta_{BL} \hat{u} \rangle_{L^2} = \\
\sum_{j=1}^S g_{p_j}(\varphi(p_j),J(p_j)) (F(p_j) - M \circ s (p_j))
\end{split}
\end{align}
for every $\varphi \in \calW$.
Moreover, equation (\ref{eq:euler-lagrange-VF}) can be reformulated as the problem of finding $(\hat{f}, \hat{h}) \in \calW \times L^2(T\Mtemp)$ that satisfies
{\small
\begin{align}\label{eq:euler-lagrange-VF_coupled}
\begin{cases}
& \langle \Delta_{BL} \hat{u},  v \rangle_{L^2} - \langle \hat{h}, v \rangle_{L^2} = 0\\
& \lambda \langle \hat{h}, \Delta_{BL} \varphi\rangle_{L^2} + \sum\limits_{j=1}^S g_{p_j}(\varphi(p_j), J(p_j)) g_{p_j}(\hat{u}(p_j),J(p_j))  = \sum\limits_{j=1}^S g_{p_j}(\varphi(p_j),J(p_j)) (F(p_j) - M \circ s (p_j))
\end{cases}
\end{align}
}
for all $(\varphi, v) \in \calW \times L^2(T\Mtemp)$. In this last reformulation, we have introduced the auxiliary function $\hat{h}$, which has been imposed to be equal, in a weak sense, to $\Delta_{BL} \hat{u}$. Now, asking the auxiliary function $v$ and the test functions $\varphi$ to be such that $v,\varphi \in \calW^1 = \{ u \in L^2(T\Mtemp) | \nabla u \in L^2(T^*\Mtemp \otimes T\Mtemp) \}$, and by exploiting the definition of the Bochner-Laplacian, we can rewrite the problem only in terms of the connection operator $\nabla$, and consequently be able to formulate it in a finite dimensional space involving only first order polynomials, as done in equation (\ref{eq:discrete-VF}).

\subsection{Vector Finite Element discretization}\label{app:VFE_discretization}

Here we introduce a linear finite element space for vector fields on a triangulated surface, where we seek for the discrete solution of the problem (\ref{eq:euler-lagrange-VF_coupled}). To this end, consider the triangulated surface $\MtempT$, approximated representation of the manifold $\Mtemp$. $\MtempT$ is not a smooth surface, so it is not even clear what the tangent space on a vertex of the triangulation is. For this reason, we use elements of computer graphics to define an interpolation basis on the triangulated surface, as done for instance in \cite{Zhang2006, Knoppel2013}.

Let now $\xi_1, \ldots,\xi_K$ be the vertices of $\MtempT$. For each vertex $\xi_k$ consider the subset of $\MtempT$ composed by the triangles adjacent to $\xi_k$, that we call here one-ring. Following the approach in \cite{Knoppel2013}, the one-ring surface is idealized by normalizing the sum of the angles incident to the vertex $\xi_k$ to add up to $2 \pi$, i.e. by `flattening' the vertex and uniformly distributing that curvature to the flat triangles of the one-ring. To the vertex $\xi_k$ they associate a unit vector basis $(e^1_k, e^2_k)$ representing a reference orientation, so that an element of the tangent vector $u_k \in T_{\xi_k}\MtempT$ will be represented by its coefficients $\vect{u}_k \in \R^2$ respect to the local basis. Then, an interpolation basis can be defined on the idealized one-ring of the vertex $\xi_k$ by parallel transporting through geodesics $(e^1_k, e^2_k)$ to the interior points of the one-ring and by scaling them with a piecewise linear function which takes value $1$ on $\xi_k$ and $0$ one the other vertices of the one-ring \cite[see][for details]{Knoppel2013}.

What is important to this work is that the outlined procedure leads to a basis of $K$ functions, whose $k$th function has support localized on the triangles adjacent to $\xi_k$, and that we denote here with the function $\bm{\psi}_k = (\psi^1_k,\psi^2_k)$, with $\psi^1_k$ and $\psi^2_k$ vector fields on $\MtempT$. For this basis functions the FE matrices $\langle \bm{\psi}_k, \bm{\psi}_{k'} \rangle_{L^2}$ and $\langle \nabla \bm{\psi}_k, \nabla \bm{\psi}_{k'} \rangle_{L^2}$ are provided.

We can finally define the FE function space $\calW_h$ to be
\begin{equation}\label{eq:VF-rep}
\calW_h = \big \{ u_h = \sum_{k=1}^K \bm{\psi}'_k \vect{u}_k | \vect{u}_k \in \R^2 \big \}.
\end{equation}

\begin{figure}[!htb]
\centering
\includegraphics[width=0.9\textwidth]{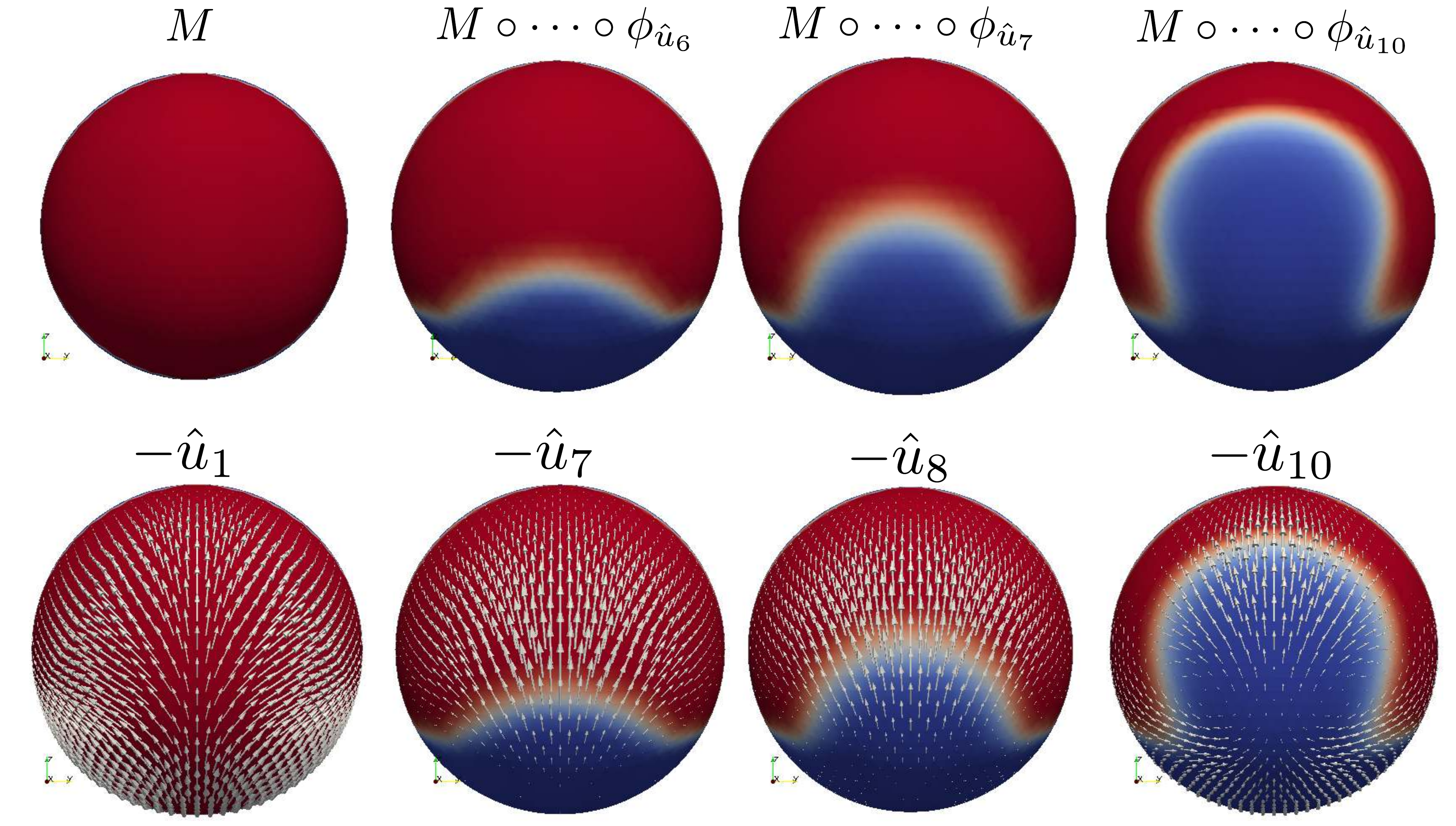}
\caption{From left to right, the estimated vector fields, and associated deformations of $M$, at 4 different iterations of the functional registration algorithm. The target is the C-shaped image $F$.}
\label{fig:fun_reg_vf}
\end{figure}

\begin{figure}[!htb]
\centering
\includegraphics[width=0.9\textwidth]{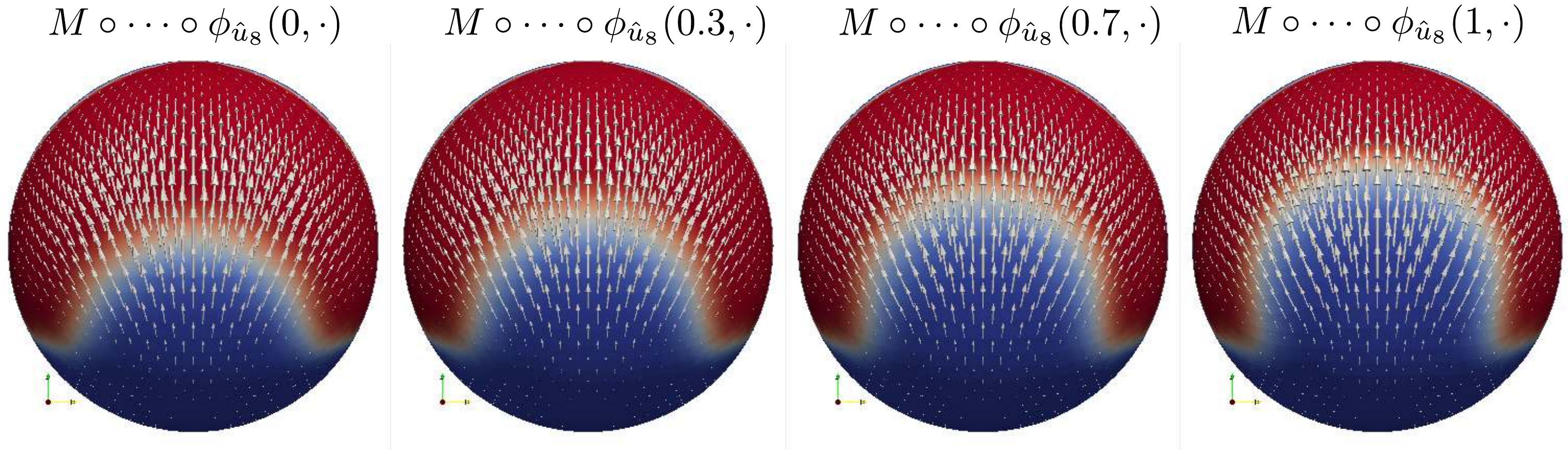}
\caption{From left to right, evolution of the flow through the ODE (\ref{eq:ODE-stationary}) for a fixed vector field. The vector field is obtained by the minimization of the linearized objective function (\ref{eq:diff-linear_app}) at the $8$th iteration.}
\label{fig:fun_reg_flow}
\end{figure}

The solution in the restricted space $\calW_h$ is finally given by the discrete approximations $\hat{u}_h, \hat{h}_h \in \calW_h$, obtained by solving
{\small
\begin{align}\label{eq:discrete-VF}
\begin{cases}
& \langle \nabla \hat{u}_h, \nabla \varphi_h \rangle_{L^2} - \langle \hat{h}_h, \varphi_h \rangle_{L^2} = 0\\
& \lambda \langle \nabla \hat{h}_h, \nabla v_h \rangle_{L^2} + \sum\limits_{j=1}^S g_{p_j}(v_h(p_j), J(p_j)) g_{p_j}(\hat{u}(p_j),J(p_j))  = \sum\limits_{j=1}^S g_{p_j}(v_h(p_j),J(p_j)) (F(p_j) - M \circ s (p_j))
\end{cases}
\end{align}
}
for all $\varphi_h, v_h \in \calW_h$.

Exploiting the representation (\ref{eq:VF-rep}) of functions in $\calW_h$ we can rewrite (\ref{eq:discrete-VF}) as a linear system as follows.  Let $\hat{\vect{u}}$ be a $2K$ vector obtained from the vectorization of the set coefficients $\{\vect{u}_i\}$. In the same way let $\hat{\vect{h}}$ be the vectorization of the coefficients of $\hat{h}_h$ in (\ref{eq:discrete-VF}). Now, introduce the $2K \times S$ matrix $\Theta_1$ and the $2K \times 2K$ matrix $\Theta_2$, such that
\begin{align*}
\vect{v}'\Theta_1 \vect{z} &= \sum\limits_{j=1}^S g_{p_j}(v_h(p_j),J(p_j)) (F(p_j) - M \circ s (p_j))\\
\vect{v}'\Theta_2 \hat{\vect{u}} &= \sum\limits_{j=1}^S g_{p_j}(v_h(p_j), J(p_j)) g_{p_j}(\hat{u}(p_j),J(p_j)),
\end{align*}
with $\vect{z}$ the vector of length $S$ such that its $j$th element is $(F(p_j) - M \circ s (p_j))$ and $\vect{v}$ the $2K$ vector obtained from the vectorization of the set coefficients of $v_h$. These sparse matrices are defined in Section~\ref{app:VF-FE_matrices}, together with the $2K \times 2K$ mass and stiffness matrices $R_0$ and $R_1$, such that
\begin{align*}
\hat{\vect{h}}' R_0 \bm{\varphi} &= \langle \hat{g}_h, \varphi_h \rangle_{L^2}\\
\hat{\vect{h}}' R_1 {\vect{v}} &= \langle \nabla \hat{g}_h, \nabla v_h \rangle_{L^2},
\end{align*}
where $\bm{\varphi}$ is a $2K$ vector obtained from the vectorization of the set coefficients $\varphi_h$.

Finally, the coefficients $\hat{\vect{u}}, \hat{\vect{h}}$, of $\hat{u}_h, \hat{h}_h$ are given by the solution of the linear system
\begin{equation}\label{eq:linear_system-VF}
	\begin{bmatrix}
		\Theta_2 & \lambda {R}_{1}\\
		\lambda {R}_{1}& -\lambda {R}_{0}
	\end{bmatrix}
	\begin{bmatrix}
		\hat{\vect{u}}\\
		\hat{\vect{h}}
	\end{bmatrix}
=
	\begin{bmatrix}
		\Theta_1 \vect{z}\\
		\vect{0}
	\end{bmatrix},
\end{equation}
where $\vect{0}$ is a $2K$ length zero-vector.

The coefficients $\{\hat{\vect{u}}_k\}$ extracted from their vectorization $\hat{\vect{u}}$ in (\ref{eq:linear_system-VF}) represent the approximated tangent vectors on the vertices $\{\xi_k\}$. They are then linearly interpolated to define a solution on $\MtempT$. This linear piecewise solution on $\MtempT$ is then used to generate a diffeomorphic transformation through the ODE (\ref{eq:ODE-stationary}), which is itself approximated with the Euler method. At each step of the Euler method the image of the solution is re-projected on $\MtempT$. Finally, the current registration is updated by composition with the newly estimated deformation as $s \la s \circ \phi_{\hat{u}}$, where $\phi_{\hat{u}}$ denotes the solution of the time $t=1$ given by the Euler method.

In Figure~\ref{fig:fun_reg_images}, we show an example of a moving image $M$, which is a semicircle indicator function, and a fixed image $F$, which is a C-shaped indicator function. They both live on the same spherical domain. This example tries to replicate the C-shaped planar registration problem, where image registration algorithms are usually tested, as for instance done in \cite{Vercauteren2009}. In Figure~\ref{fig:fun_reg_vf} we show the vector fields estimated at four different iterations of the Algorithm~\ref{alg:fun-reg}. While in Figure~\ref{fig:fun_reg_flow}, for one particular iteration, we show the evolution of the flow generated by the ODE (\ref{eq:ODE-stationary}). In this specific example, the domain is chosen to be spherical for visualization purposes, however it can be any smooth two-dimensional manifold, as for instance, in Section~\ref*{sec:application}. The performances of the algorithm, with these synthetic data, are excellent. In fact, only 12 iterations are necessary to register the semicircled indicator function to the C-shaped indicator function.

Finally, it could be argued that being the proposed approximation of the vector field $\hat{u}$ only piecewise linear, and not of higher regularity, this could lead to deformations that are not diffeomorphic. However, the use of reasonably fine triangulated meshes $\MtempT$ should solve the problem. After all, in practice, even for higher regularity vector fields, the computer resolution of the ODE relies on a finite number of sampled values from the vector field, and thus on a non smooth vector field.

\subsection{Finite element matrices}\label{app:VF-FE_matrices}
Assume, for simplicity, that the points $\{p_j\}$ coincide with the nodes $\{\xi_k: 1,\ldots,K\}$ of the mesh $\MtempT$. The non-zero entries of the matrices $\Theta_1$  and $\Theta_2$ are
\begin{align*}
&\{\Theta_1\}_{2k,k} = g_{\xi_k}(J(\xi_k),e_1^k), \\
&\{\Theta_1\}_{2k+1,k} = g_{\xi_k}(J(\xi_k),e_2^k)
\end{align*}
and
\begin{align*}
&\{\Theta_2\}_{2k,2k} = g_{\xi_k}^2(J(\xi_k),e_1^k), \qquad
\{\Theta_2\}_{2k,2k+1} = g_{\xi_k}(J(\xi_k),e_1^k)g_{\xi_k}(J(\xi_k),e_2^k),\\
&\{\Theta_2\}_{2k+1,2k} = g_{\xi_k}(J(\xi_k),e_1^k)g_{\xi_k}(J(\xi_k),e_2^k), \qquad
\{\Theta_2\}_{2k+1,2k+1} = g_{\xi_k}^2(J(\xi_k),e_2^k)
\end{align*}
with the matrices indexed from zero and $k = 0, \ldots, K-1$. The computation of the entries $g_{\xi_k}(J(\xi_k),e_1^k)$ can be performed by representing the tangent vectors $J(\xi_k)$ and $e_1^k$ as vectors in $\R^3$ and computing the $\R^3$ Euclidean scalar product between them, as in fact the manifold $\Mtemp$, and its associated triangulated mesh $\MtempT$, are embedded in $\R^3$. The entries of the $2K \times 2K$ matrices ${R}_0$ and ${R}_1$ in (\ref{eq:discrete-VF}) are computed in \cite[][Section 6.1.1]{Knoppel2013}, for the purpose of computing eigen-vectors of the Bochner-Laplacian operator.

\subsection{Boundary Conditions}\label{app:BC}
The deformations generated by the functional registration algorithm are by definition constrained to be maps with their image on the template surface, since the ODE (\ref{eq:ODE-stationary}) is defined on the manifold itself. However, if the template is a manifold with a boundary, as in the simulations performed in Section~\ref{sec:simulations}, the vector might generate deformations that transport the functions outside the boundary. This can be avoided by imposing homogeneous Dirichlet boundary conditions on the estimated vector field. Dirichlet boundary conditions can be implemented in different ways. Here, we opt for applying them after the linear system (\ref{eq:linear_system-VF}) has been built. In particular given a boundary node $k$, we add a large constant $M$ to the entries $2k,2k$ and $2k+1,2k+1$ of the left hand side matrix and set to $0$ the entries $2k$ and $2k+1$ of the right hand side vector. As a consequence, the vector fields estimated from the modified linear system will smoothly vanish as approaching the boundary.

\section{Further Simulations}\label{app:further_simulations}
As previously mentioned, the functional registration algorithm introduced in Section~\ref*{sec:functional_registration_brief} is not the only option to account for functional information in the registration process. Here we compare our methodology to the joint functional and geometric registration algorithm proposed in \cite{Charon2014}, where the shape similarity functional (\ref{eq:empirical}) is extended to include a functional similarity term.

Suppose now that the template mesh $\MtempT$ is equipped with a functional object $\mu^\calT:\MtempT \ra \R$, which in first instance can be the cross-sectional mean of the functions $\hat{X}_i$ estimated after the geometric registration described in Section~\ref*{sec:geometric_registration_brief}. We briefly recall the notation in Section~\ref*{sec:geometric_registration_brief}, introduced to define (\ref{eq:empirical}). We define $K_{\calZ}: \R^3 \times \R^3 \ra \R^{3 \times 3}$ to be a Gaussian isotropic kernel of variance $\sigma^2_\calZ$, i.e. $K_{\calZ}(x,y)= \exp(-\| x-y \|_2^2/ (2\sigma_{\calZ}^2)) \Id_{3 \times 3}$, with $\Id_{3 \times 3}$ denoting a $3 \times 3$ identity matrix. Additionally, we introduce a scalar Gaussian kernel $K_{\calF}: \R \times \R \ra \R$ of the form $K_{\calF}(x,y)= \exp(-(x-y)^2/ (2\sigma_{\calF}^2))$.

\begin{figure}[!htb]
\centering
\includegraphics[width=0.85\textwidth]{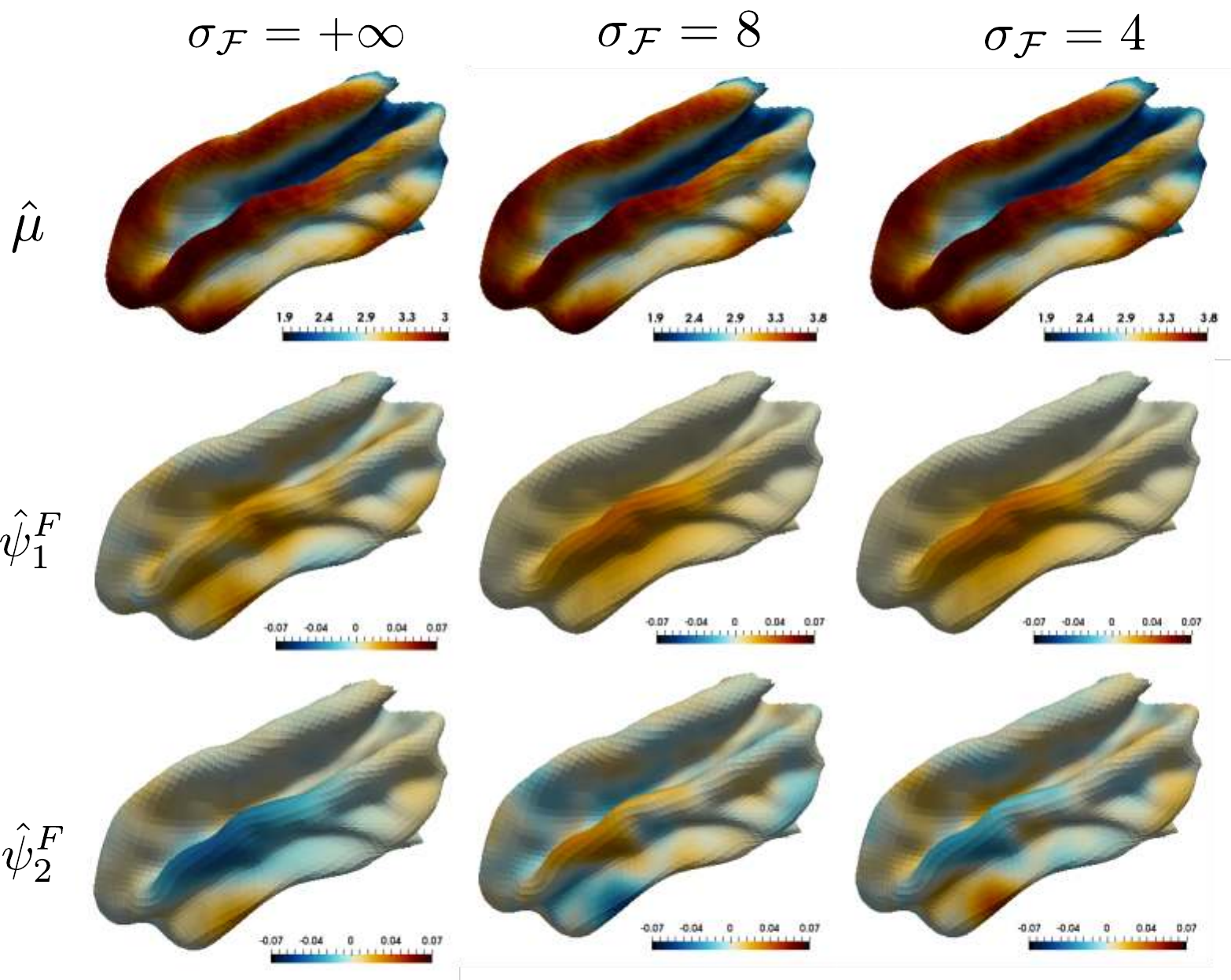}
\caption[]{From left to right, the mean and first two functional PC functions estimates of $\{\hat{X}_i\}$, estimated by using the registration maps computed by solving (\ref{eq:minimization_geo}) with the extended matching function in (\ref{eq:empirical-fcurr}), for different choices of $\sigma_\calF$.}
\label{fig:fPCA_fshapes}
\end{figure}

Moreover, we denote with $c(l)$ and $\eta(l)$, respectively, the center point and the normal vector of the $l$th triangle of the mesh $\varphi_{v_i} \circ \M_0^\calT$. We denote with $c_i(q)$ and $\eta_i(q)$, respectively, the center point and the normal vector of the $q$th triangle of the mesh $\calM^\calT_i$. Additionally, we introduce $y(l)$, denoting the functional value $\mu^\calT$, associated to the mesh $\varphi_{v_i} \circ \M_0^\calT$, at the center point of the $l$th triangle. We denote with $y_i(q)$ the functional value associated to the $i$th \FoSs\ at the center point of the $q$th triangle of the mesh $\M_i$.

Let the triangles of the mesh $\varphi_{v_i} \circ \M_0^\calT$ be indexed by $l$ and $g$ and the triangles in $\calM^\calT_i$ be indexed by $q$ and $r$. The shape similarity functional (\ref{eq:empirical}) can be extended to include functional informations as follows \citep{Charon2014}.
\begin{align}\label{eq:empirical-fcurr}
\begin{split}
D^2 \big((\varphi_{v_i} &\circ \calM^\calT_0, \mu^\calT \circ \varphi^{-1}_{v_i}), (\calM^\calT_i, Y_i^\calT) \big) =\\
&\sum_l \sum_g K_{\calF}(y(l),y(g)) K_{\calZ} (c(l),c(g)) \eta(l) \cdot \eta(g)\\
&-2 \sum_l \sum_q K_{\calF}(y(l),y_i(q)) K_{\calZ}(c(l),c_i(q)) \eta(l) \cdot \eta_i(q)\\
&+ \sum_q \sum_r K_{\calF}(y_i(q),y_i(r)) K_{\calZ}(c_i(q),c_i(r)) \eta_i(q) \cdot \eta_i(r),
\end{split}
\end{align}
with $\cdot$ denoting the scalar product in $\R^3$.
Each term now, measures not only differences in geometry but also differences in the functional values between the template and the target \FoS.

Subsequently, given the \FoSs\ $\{(\M_i^\calT,Y_i^\calT)\}$ generated as described in Section~\ref{sec:simulations}, we perform the landmark-free geometric registration by minimizing the objective function in (\ref{eq:minimization_geo}), with the shape similarity functional (\ref{eq:empirical}), which is equivalent to the similarity functional (\ref{eq:empirical-fcurr}) with $\sigma_\calF = +\infty$. Thanks to the estimated registration maps we can estimate the functions $\{\hat{X}_i\}$ and compute the cross-sectional mean function $\mu^\calT$. Subsequently a second registration step can be performed, by minimizing the objective function in (\ref{eq:minimization_geo}) but this time with the similarity functional (\ref{eq:empirical-fcurr}). We have performed this for different choice of $\sigma_\calF$. The smaller $\sigma_\calF$, the more we are weighting the functional matching term as opposed to the geometric matching term.

In Figure~\ref{fig:fPCA_fshapes} we show the results of the fPCA applied to the functions $\{\hat{X}_i\}$ for different choices of $\sigma_\calF$. These need to be compared with the results in Figure~\ref{fig:fPCA}, obtained by applying the iterative functional registration algorithm in Section~\ref*{sec:functional_registration_brief}. On the left panel of Figure~\ref{fig:fPCA_fshapes} we can see the mean and first two PC functions estimated when functional information is ignored, which coincide with the one showed on the left panel of Figure~\ref{fig:fPCA}, as they are computed in the same way. On the other two panels of Figure~\ref{fig:fPCA_fshapes} we can see the mean and first two PC functions estimated when functional information is introduced. As we can see the estimated first PC function resembles the true underlying first PC function, but some fictitious variability is left on the second estimated PC function.

Trying to further decrease $\sigma_\calF$, to remove the residual fictitious variability, resulted in estimated registration maps failing to bring the template in geometric correspondence to the target surface. Such problem has been the limiting factor in successfully applying the same method to the data in the real application, where the differences in geometries between the template and the target \FoSs\ are much bigger. In fact, this is one of the motivations underlying the introduction of the functional registration algorithm in Section~\ref*{sec:functional_registration_brief}, where the `moving' functions are instead `constrained' to lie in the predefined geometry.

\bibliographystyle{abbrvnat_brief}

\bibliography{Bibliography} 

\end{document}